\documentclass[11pt,a4paper]{article}

\usepackage{geometry}
\usepackage{ulem}%\geometry{left=32mm,right=32mm,top=25mm,bottom=25mm}
\usepackage{microtype}           % Better typography
\usepackage{makecell}
\usepackage{amsmath,amssymb,amsthm}
\usepackage{mathtools}
\usepackage{physics}
\usepackage{bm}
\usepackage{graphicx}
\usepackage{subfigure}
\usepackage{caption}
\usepackage{mathrsfs}
\usepackage{amsthm}
\usepackage[colorlinks=true,linkcolor=black,citecolor=blue,urlcolor=blue,bookmarks]{hyperref}

\usepackage{authblk}

\usepackage{comment}
\usepackage{appendix}
\usepackage{tikz}
\usetikzlibrary{arrows.meta}

\def \be {\begin{equation}}
\def \ee {\end{equation}}
\def \ba {\begin{array}}
\def \ea {\end{array}}
\def \bea{\begin{eqnarray}}
\def \eea{\end{eqnarray}}
\def \nn {\nonumber}

\title{Entanglement of General Subregions in Time-Dependent States}

\author{Wu-zhong Guo${}^{1,}$\footnote{wuzhong@hust.edu.cn},
\quad Song He${}^{2,}$\footnote{hesong@nbu.edu.cn},
\quad Tao Liu${}^{1,}$\footnote{liu\_tao\_@hust.edu.cn}
\\

\small ${}^{1}$School of Physics, Huazhong University of Science and Technology,\\
\small Luoyu Road 1037, Wuhan, Hubei 430074, China
\\
\small ${}^{2}$Institute of Fundamental Physics and Quantum Technology \& School of Physical Science and Technology, Ningbo University\\
}
\date{\today}

\begin{document}

\maketitle

\begin{abstract}

%We present a unified real-time framework for computing Rényi and entanglement entropies of arbitrary spacetime intervals in time-dependent states of (1+1)-dimensional conformal field theories. Using the spacetime density matrix and a Schwinger--Keldysh replica construction, we obtain analytic expressions valid for both spacelike and timelike separations. Applying the method to global quenches prepared by boundary states and to local quenches generated by operator insertions, we find that spacelike intervals display linear growth followed by saturation, while timelike intervals yield time-independent entanglement with a universal imaginary term ${i\pi c\over 6}$. All regimes admit a single quasiparticle interpretation: entanglement is produced precisely when one worldline of each pair intersects the interval. We further show that the linear relation connecting spacelike and timelike entanglement persists in both quench models, indicating a broader universality than previously known. Our results provide a systematic characterization of spacetime entanglement in real-time quantum dynamics.

We develop a unified framework for computing Rényi and entanglement entropies of arbitrary spacetime intervals in time-dependent states of $(1+1)$-dimensional conformal field theories. By combining the spacetime density matrix formalism with the replica method, we show that entanglement entropy is well defined for both spacelike and timelike separations. Applying this framework to global quenches prepared by boundary states and to local quenches generated by operator insertions, we obtain analytic expressions for the entanglement entropy in general spacetime configurations. The results reveal qualitative differences between spacelike and timelike intervals: the timelike entanglement entropy is time-independent in the global quench model, depends solely on the temporal separation, and universally exhibits a constant imaginary contribution. These features are naturally explained by a generalized quasiparticle picture in which entanglement is produced precisely when one  worldline of each quasiparticle pair intersects the interval. Furthermore, we demonstrate that the linear sum rule relating time- and spacelike entanglement persists in both global and local quenches, indicating a broader universality of spacetime entanglement in real-time quantum dynamics.

%We develop a unified real-time framework for computing Rényi and entanglement entropies of arbitrary spacetime intervals in time-dependent states of $(1+1)$-dimensional conformal field theories. By combining the spacetime density matrix formalism with the Schwinger--Keldysh real-time replica method, we show that entanglement entropy is well defined for both spacelike and timelike separations. Applying this framework to global quenches prepared by boundary states and to local quenches generated by operator insertions, we obtain analytic expressions for the entanglement entropy in general spacetime configurations. The results reveal qualitative differences between spacelike and timelike intervals: the timelike entanglement entropy is time-independent, depends solely on the temporal separation, and universally exhibits a constant imaginary contribution. These features are naturally explained by a generalized quasiparticle picture in which entanglement is produced precisely when one worldline of each quasiparticle pair intersects the interval. Furthermore, we demonstrate that the linear sum rule relating time- and spacelike entanglement persists in both global and local quenches, indicating a broader universality of spacetime entanglement in real-time quantum dynamics.

\end{abstract}
\newpage

\tableofcontents

\section{Introduction}
Entanglement entropy (EE) has emerged as a central tool for probing the fundamental structure of quantum many-body systems and quantum field theories (QFTs) \cite{Amico:2007ag,Calabrese:2009qy,Zeng:2015pxf}. Its holographic formulation \cite{Ryu:2006bv,Hubeny:2007xt} has led to deep insights into the geometric nature of gravity and the quantum structure of black holes 
\cite{VanRaamsdonk:2010pw,Swingle:2009bg,Maldacena:2013xja,Almheiri:2014lwa,
Jafferis:2015del,Dong:2016eik,Engelhardt:2014gca,Penington:2019npb,
Almheiri:2019psf,Almheiri:2020cfm}; see also \cite{Takayanagi:2025ula} for a recent review. At a fundamental level, EE provides a non-perturbative probe of the quantum state, capturing information inaccessible to local observables. Its significance has also
been clarified from an algebraic perspective, where entanglement is understood in terms of operator algebras and modular structure
\cite{Witten:2018zxz,Witten:2021unn,Chandrasekaran:2022cip,
Chandrasekaran:2022eqq,Guo:2018lqq}.

EE is a powerful probe of the dynamical properties of quantum
many-body systems. Its evolution following quantum quenches has yielded deep insight
into non-equilibrium dynamics in isolated systems. In $(1+1)$-dimensional conformal
field theories (CFTs), the imaginary-time path-integral approach provides a systematic
framework for analyzing the growth and saturation of EE after global and local quenches
\cite{Calabrese:2005in,Calabrese:2006rx,Calabrese:2007rg,Calabrese:2007mtj,
Nozaki:2014hna,Nozaki:2014uaa,Caputa:2014vaa,He:2014mwa,Asplund:2014coa,
Guo:2015uwa,He:2017lrg}; see also the review \cite{Calabrese:2016xau}. This framework
has since been widely applied to studies of entanglement dynamics and thermalization.

Most existing work, however, focuses on spatial subregions taken at equal time. This
restriction is natural in the Euclidean path-integral preparation of the initial density
matrix $\rho_0$, followed by unitary evolution to $\rho(t)$. In this setting, dynamical
information is extracted by comparing the EE of spatial subregions on a fixed Cauchy
slice. From a relativistic perspective, however, entanglement is intrinsically a spacetime
notion, and a complete characterization requires extending EE to general spacetime
subregions, including causally connected ones. An early attempt in this direction is the
proposal of timelike EE, defined by formally treating time as a spatial direction
\cite{Doi:2022iyj}.

Recently, it was shown that the EE for causally connected subregions can be consistently
defined and computed \cite{Milekhin:2025ycm}. This construction finds a natural and
systematic formulation in the spacetime density matrix formalism
\cite{Guo:2025dtq}, which generalizes the standard density matrix on a single Cauchy
surface to one defined on multiple Cauchy surfaces, thereby encoding timelike
correlations.  Within this framework, the EE of causally connected subregions can be
evaluated using the Schwinger--Keldysh path integral and the real-time replica method
\cite{Milekhin:2025ycm,Gong:2025pnu,Guo:2025dtq,Dong:2016hjy,
Colin-Ellerin:2020mva,Colin-Ellerin:2021jev}. In appropriate limits, this approach
reproduces the timelike EE proposed in \cite{Doi:2022iyj}.  Closely related ideas have also been explored from a quantum information perspective,
where temporal correlations are encoded in generalized density matrices; see, e.g.,
\cite{Fitzsimons:2013gga,Buscemi:2013xlk,LS,Horsman:2017,Fullwood:2022rjd,
Parzygnat:2022pax,Lie:2024kbl,Fullwood:2025zxu,Lie:2025pst,Diaz:2020dfe,
Diaz:2021snw,Diaz:2023npx,Diaz:2025aqe}.

Previous studies of timelike EE have largely been restricted to static states, such as the
vacuum or thermal ensembles. For timelike EE to constitute a meaningful extension of
entanglement, it must be well defined in generic, time-dependent states. One of the
primary motivations of this work is to address this question. As we demonstrate in
Sec.~\ref{Section_set_up}, spacelike and timelike subregions can be treated on equal
footing within the spacetime density matrix formalism, and both Rényi entropy and EE
for arbitrary spacetime subregions can be computed using the real-time replica method
in quantum field theory.

%In this work, we extend the study of entanglement entropy to general spacelike and timelike intervals, offering both analytical results and physical interpretations.

More precisely, we systematically analyze the behavior of EE for general spacetime
subregions in two well-studied quantum quench models: the global quench
\cite{Calabrese:2006rx,Calabrese:2007rg} and the local quench
\cite{He:2014mwa}. We develop a unified method to compute Rényi entropies and EE
for arbitrary spacetime intervals in $(1+1)$-dimensional CFTs, and use it to derive
explicit results in various configurations, thereby classifying their dynamical behavior.

A central observation in both quench models is that the entanglement dynamics admits
a quasiparticle interpretation
\cite{Calabrese:2005in,Calabrese:2006rx,Calabrese:2007rg,Alba:2017ekd,Bastianello:2018fvl}.
We show that this picture extends naturally to EE associated with general spacetime
intervals, providing a unified interpretation of early-time growth, linear scaling, and
saturation in terms of the trajectories of entangled quasiparticles crossing the
entangling region. In this sense, our results furnish a natural generalization of the
quasiparticle picture to arbitrary spacetime subregions.

It was shown in \cite{Guo:2024lrr} that a simple sum rule relates time- and spacelike EE
in certain states of $(1+1)$-dimensional CFTs. For more general states, one expects
controlled modifications, which can be constructed perturbatively
\cite{Xu:2024yvf}, and the relation has been further extended to higher dimensions
\cite{Guo:2025pru}. Although ultimately rooted in causality constraints in QFTs, it is
remarkable that EE obeys such a simple linear structure. This sum rule highlights the
intrinsic connection between time- and spacelike entanglement and, through its
holographic interpretation, implies an exact correspondence between black hole
interior and exterior regions \cite{Guo:2025pru}, suggesting the possibility of
reconstructing the interior from exterior data \cite{Guo:2025mwp}.

For time-dependent states, verifying the validity of this sum rule is highly non-trivial.
A main result of this work is to show that the sum rule continues to hold in both the
global and local quench models considered here, providing evidence that it applies
to a much broader class of dynamical CFT states.

The remainder of this paper is organized as follows. In
Sec.~\ref{Section_set_up}, we review the spacetime density matrix formalism and the
real-time replica method for general spacetime subregions. In Sec.~\ref{sec:3}, we
apply this framework to global quenches, deriving the EE for equal-time, spacelike,
and timelike intervals and interpreting the results in a unified quasiparticle picture.
Sec.~\ref{sec:4} analyzes local quenches induced by operator insertions, including
explicit examples in free and rational CFTs, and extends the quasiparticle
interpretation to general spacetime intervals. In Sec.~\ref{sec:5}, we examine the
relation between time- and spacelike EE and demonstrate that the linear sum rule
remains valid in both quench models. We conclude in Sec.~\ref{sec:6} with a discussion
of implications and possible extensions. Additional technical details and conventions
are collected in the appendices.

%\textcolor{red}{
%1. previous studies on entanglement: timeslice subregions, quantum quench, quasi-particle picture...\\
%2. In this paper, entanglement for general subregions in spacetime, specificly, causally connected subregions- timelike entanglement. 
%3. previous studies on timelike entanglement most focus on static state, such as vacuum state and thermal state etc.  In this paper we consider the timelike entanglement for the dynamical states.
%4. Relation between time- and spacelike entanglement entropy.
%}

\section{Entanglement for general subregions in spacetime}\label{Section_set_up}

%\textcolor{red}{
%1. (Reduced) Spacetime density matrix and SK path integral\\
%2. Timelike entanglement by spacetime density matrix\\
%3. Quantum quench: global state and local quench state\\
%4. Timelike entanglment for dyanmaical states\\
%5 How to calculate timelike EE (Wick rotation)
%}
\subsection{Spacetime density matrix formalism}

In this section, we define entanglement for arbitrary spacetime subregions using the
spacetime density matrix formalism. The spacetime density matrix encodes correlation
information across multiple Cauchy surfaces and can be viewed as a natural generalization
of the conventional density matrix defined on a single Cauchy surface. In appropriate
limits, it reduces to the standard density matrix.

Consider two Cauchy surfaces $C_0: t = t_0$ and $C_1: t = t_1$, with $t_1 > t_0 > 0$, and
an initial density matrix $\rho_0$. We associate Hilbert spaces $\mathcal{H}_0$ and
$\mathcal{H}_1$ with $C_0$ and $C_1$, respectively. One may then evaluate the Wightman
correlation functions
\bea
\langle \mathcal{O}_0(t_0)\mathcal{O}_1(t_1)\rangle_{\rho_0}
:= \Tr\!\left(\rho_0\,\mathcal{O}_0(t_0)\mathcal{O}_1(t_1)\right).
\eea
It can be shown that there exists an operator
$T_{C_0C_1}:\mathcal{H}_0\otimes\mathcal{H}_1\to\mathcal{H}_0\otimes\mathcal{H}_1$
\cite{Milekhin:2025ycm,Guo:2025dtq} such that
\bea
\Tr\!\left(T_{C_0C_1}\mathcal{O}_0\mathcal{O}_1\right)
= \langle \mathcal{O}_0(t_0)\mathcal{O}_1(t_1)\rangle_{\rho_0}.
\eea
Here we fix the Wightman function to be anti–time-ordered. One could also study time-ordered or even out-of-time-ordered correlators (OTOC) \cite{Das:2025fcd}.
Choosing bases $\{|i\rangle\}$ for $\mathcal{H}_0$ and $\mathcal{H}_1$, the operator
$T_{C_0C_1}$ can be written explicitly as
\bea\label{Spacetime_DM_formula}
T_{C_0C_1}
= \langle l|U(t_1,t_0)\rho_0|i\rangle
  \langle j|U^\dagger(t_1,t_0)|k\rangle\,
  |j\rangle\langle i|\otimes|l\rangle\langle k|,
\eea
where summation over repeated indices is implicit, and
$U(t_1,t_0)=e^{-iH(t_1-t_0)}$ is the time-evolution operator. The operator
$T_{C_0C_1}$ is referred to as the spacetime density matrix. Further details on its
derivation, generalizations, and properties can be found in \cite{Guo:2025dtq}.

The focus of this work is entanglement between subregions defined on different Cauchy
surfaces. Let $A_0\subset C_0$ and $A_1\subset C_1$ be such subregions. The reduced
spacetime density matrix associated with $A_0$ and $A_1$ is defined by tracing out the
complementary degrees of freedom,
\bea\label{Reduced_Spacetime_DM}
T_{A_0A_1}:=\Tr_{\bar A_0\bar A_1} T_{C_0C_1},
\eea
where $\bar A_0$ and $\bar A_1$ denote the complements of $A_0$ and $A_1$, respectively.
The operator $T_{A_0A_1}$ acts on the Hilbert space
$\mathcal{H}_{A_0}\otimes\mathcal{H}_{A_1}$.

Given $T_{A_0A_1}$, one can define entropy-related quantities, such as the Rényi entropy
for integer $n>1$,
\bea\label{Renyi_definition}
S_n(T_{A_0A_1})
:= \frac{1}{1-n}\log\Tr\!\left(T_{A_0A_1}^n\right),
\eea
with the von Neumann entropy obtained in the limit
$S(T_{A_0A_1})=\lim_{n\to1}S_n(T_{A_0A_1})$.

In QFTs, both $T_{C_0C_1}$ and $T_{A_0A_1}$ admit representations in
terms of the Schwinger--Keldysh path integral. This allows the Rényi entropy
\eqref{Renyi_definition} to be evaluated using the replica method within a real-time
path-integral framework, in close analogy with the Euclidean formulation. This
construction provides a natural framework for defining timelike entanglement, namely,
entanglement between causally connected subregions. While previous studies have focused
primarily on static states, such as the vacuum or thermal states, in this work we show
that the same formalism applies straightforwardly to time-dependent states.

\subsection{General set-up}

In this section, we show how to construct the (reduced) spacetime density matrix for time-dependent states. One may prepare a state $|\Psi_0\rangle$ at time $t=0$. At time $t_0 > 0$, the state is given by
\bea
|\Psi(t_0)\rangle=U(t_0,0)|\Psi_0\rangle.
\eea
Now with the state $|\Psi(t_0)\rangle$, or density matrix $\rho_0=|\Psi(t_0)\rangle \langle \Psi(t_0)|$, one could construct the spacetime density matrix for the Cauchy surfaces $C_0$ and $C_1$ by using the formula (\ref{Spacetime_DM_formula}). The result is
\bea
\tilde{T}_{C_0C_1}=\langle l|U(t_1,0)|\Psi_0\rangle \langle \Psi_0| U(t_0,0)^\dagger |i\rangle \langle j|U(t_1,t_0)^\dagger|k\rangle  |j\rangle \langle i| \otimes |l\rangle \langle k|,
\eea
where we have used the fact $U(t_1,t_0)U(t_0,0)=U(t_1,0)$. One can also obtain its Hermitian conjugate, $\tilde{T}_{C_0C_1}^\dagger$.

\begin{figure}[htbp]
    \centering
    \subfigure{
        \includegraphics[width=0.38\textwidth]{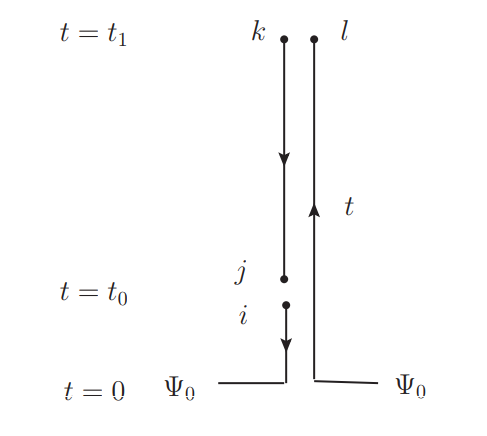}}
    \subfigure{
        \includegraphics[width=0.38\textwidth]{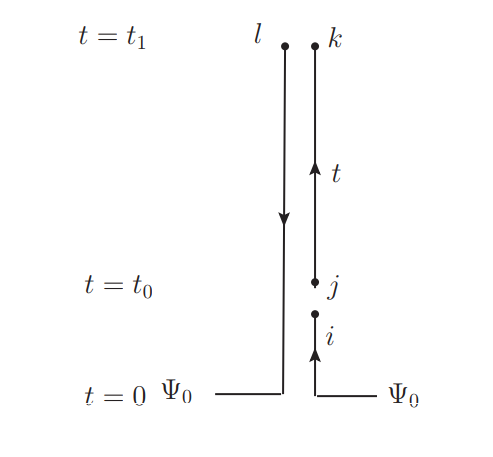}}
        \caption{Schwinger–Keldysh representation of the spacetime density matrix $\tilde{T}_{C_0C_1}$ (left) and its Hermitian conjugate $\tilde{T}_{C_0C_1}^\dagger$ (right). The initial state at time $t=0$ is denoted by $\Psi_0$, and $i,j,k,l$ denote the boundary conditions. The arrows indicate the direction of the time evolution operator.}\label{Fig_SK_T01}
\end{figure}

We can now represent the operators $\tilde{T}_{C_0C_1}$ and $\tilde{T}_{C_0C_1}^\dagger$ in terms of the Schwinger--Keldysh path integral, as illustrated in Fig.~\ref{Fig_SK_T01}.

We can check the simple properties satisfied by $\tilde{T}_{C_0C_1}$. If making partial trace on $C_0$, we would have $Tr_{C_0} \tilde{T}_{C_0C_1}=\rho(t_1):=|\Psi(t_1)\rangle \langle \Psi(t_1)|$. While, if making partial trace on $C_1$, we would have $Tr_{C_1} \tilde{T}_{C_0C_1}=\rho(t_0):=|\Psi(t_0)\rangle \langle \Psi(t_0)|$. One could check these two properties by using the path integral representation Fig.\ref{Fig_SK_T01}\footnote{In the path integral representation, making a trace on $C_0$ means gluing the boundary conditions $i$ and $j$, while making a trace on $C_1$ means gluing the boundary conditions $k$ and $l$. }.  Thus, the spacetime density matrix $\tilde{T}_{C_0C_1}$ can be seen as a natural generalization of the time-dependent density matrix $\rho(t)$.

In QFTs, $A_0$ and $A_1$ are taken to be spatial subregions. In this
work, we focus on $(1+1)$-dimensional QFTs, where $A_0$ and $A_1$ are chosen as spatial
intervals. Motivated by computational simplicity,\footnote{With this choice, the Rényi
entropy reduces to a two-point function of twist operators, allowing for explicit
analytical results in the models considered.} we take
\bea
A_0 = (-\infty,x_0), \qquad A_1 = (x_1,\infty),
\eea
where $x_0$ and $x_1$ denote the spatial endpoints of $A_0$ and $A_1$, respectively. We
denote the interval between the spacetime points $(t_0,x_0)$ and $(t_1,x_1)$ by
$A(t_0,x_0;t_1,x_1)$. Depending on the relative separation of these points, the interval
can be spacelike or timelike. Both cases are treated within a unified framework in our
setup.

\begin{figure}[htbp]
  \centering
  \includegraphics[width=0.9\textwidth]{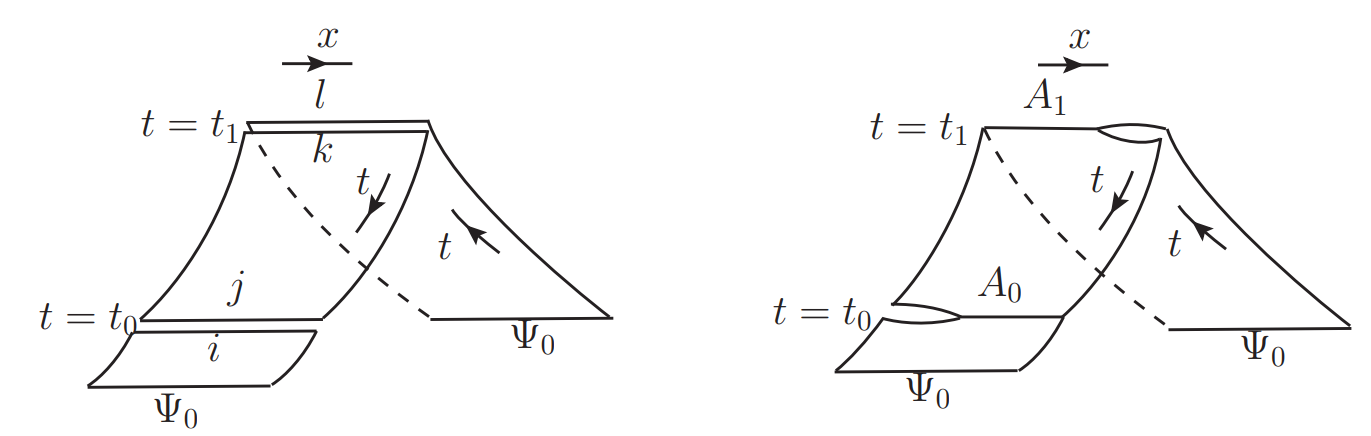}
  \caption{(Left) Schwinger--Keldysh representation of the spacetime density matrix
  $\tilde{T}_{C_0C_1}$, including the spatial direction $x$. The arrow along the time axis
  indicates the direction of time evolution. (Right) Representation of the reduced
  spacetime density matrix $\tilde{T}_{A_0A_1}$. The degrees of freedom on $\bar A_0$ and
  $\bar A_1$ are glued together, while cuts remain along $A_0$ and $A_1$.}
  \label{Fig_SK_T01_reduced_quench}
\end{figure}

The reduced spacetime density matrix $\tilde{T}_{A_0A_1}$ admits a representation in
terms of the Schwinger--Keldysh path integral, as illustrated in
Fig.~\ref{Fig_SK_T01_reduced_quench}.

Using this path-integral representation, the moments of $\tilde{T}_{A_0A_1}$ can be
computed via the replica method, in close analogy with the Euclidean formulation. For
discussions of the real-time replica method, see
\cite{Dong:2016hjy,Colin-Ellerin:2020mva,Colin-Ellerin:2021jev}. One finds
\bea\label{n_moments}
\Tr\!\left(\tilde{T}_{A_0A_1}^n\right)
= \langle\Psi_0^n|\tilde{\sigma}_n(t_0,x_0)\sigma_n(t_1,x_1)|\Psi_0^n\rangle,
\eea
where $|\Psi_0^n\rangle$ denotes the $n$-fold replicated state, and $\sigma_n$ and
$\tilde{\sigma}_n$ are the twist and anti-twist operators, with conformal dimensions
$h_n=\bar h_n=\frac{c}{24}\!\left(n+\frac{1}{n}\right)$. The Rényi entropy is then
obtained using definition~\eqref{Renyi_definition}.

Thus, the computation of Rényi entropy reduces to the evaluation of Lorentzian
correlation functions of twist operators in the state $|\Psi_0^n\rangle$. The formalism
applies to both spacelike and timelike intervals $A(t_0,x_0;t_1,x_1)$, although, as we
will show, their entanglement properties exhibit qualitatively different behavior.

In the remainder of this work, we focus on two time-dependent states commonly used in
the study of quantum quenches: the global quench and the local quench in
$(1+1)$-dimensional CFTs. Their constructions are reviewed in
Secs.~\ref{sec:3} and~\ref{sec:4}. In suitable limits, analytic results can be obtained.
Lorentzian correlators are evaluated via analytic continuation of Euclidean correlators
\cite{Hartman:2015lfa}. In particular, for
$\Tr(\tilde{T}_{A_0A_1}^n)$ in~\eqref{n_moments}, we perform the continuation
\bea\label{analytical_continuation}
\Tr\!\left(\tilde{T}_{A_0A_1}^n\right)
= \langle\Psi_0^n|\tilde{\sigma}_n(\tau_0,x_0)\sigma_n(\tau_1,x_1)|\Psi_0^n\rangle
\Big|_{\tau_0\to it_0+\epsilon_0,\;\tau_1\to it_1+\epsilon_1},
\eea
with $\epsilon_0>\epsilon_1>0$.\footnote{This $i\epsilon$-prescription fixes the operator
ordering and plays a crucial role in obtaining the imaginary part of the timelike
EE, as discussed in Sec.~\ref{sec:5}.} For the states considered in
this paper, the Euclidean correlators can be evaluated explicitly in appropriate limits,
allowing the Lorentzian results to be obtained by analytic continuation.

EE has been extensively used to probe non-equilibrium dynamics; see
\cite{Calabrese:2016xau} for a review. Typically, one prepares an initial state
$|\Psi_0\rangle$ at $t=0$ and studies the time-evolved state
$|\Psi(t)\rangle=U(t,0)|\Psi_0\rangle$. The spacetime density matrix $T_{C_0C_1}$ provides
a natural generalization of the single-time density matrix $\rho(t)$ to multiple time
slices. In the special case $t_0=t_1$, Eq.~\eqref{n_moments} reduces to a correlator on a
single time slice, reproducing the standard EE of $|\Psi(t)\rangle$.
Our results therefore recover known results in this limit, while extending them to
general spacetime intervals.

\section{EE in global quench model} 
\label{sec:3}
% A global quench describes a process in which the Hamiltonian of the system undergoes a sudden and spatially uniform change across the entire space, causing the system to evolve rapidly from a near-ground-state configuration to a nonequilibrium state. Typically, the system is initially prepared in the ground state of a pre-quench Hamiltonian $H_0$, and at time $t_0 = 0$, the Hamiltonian is abruptly changed to $H$. Thereafter, the system evolves unitarily under the new Hamiltonian as
% behavior is simple in the limits $\eta\to 0$ and $\eta\to 1$. 
% When $\eta\to 1$, the bulk identity block dominates and
% $F(1)=1.$ For $\eta\ll 1$, the short-distance bulk--boundary OPE yields
% $ F(\eta)\simeq (A_b^{\Phi})^{2}\,\eta^{x_b},$ 
% where $A_b^{\Phi}$ is the bulk-to-boundary OPE coefficient and $x_b$
% is the scaling dimension of the leading boundary operator.  
% For twist operators, the leading boundary operator is the identity, 
% so $x_b=0$, and the expression simplifies to
% $F(\eta)\simeq (A_b^{\Phi})^{2}$. For further details, see Ref.~\cite{Calabrese:2016xau}.

A global quench describes a process in which the Hamiltonian of the system undergoes a sudden and spatially uniform change across the entire space, causing the system to evolve rapidly from a near-ground-state configuration to a nonequilibrium state. Typically, the system is initially prepared in the ground state of a pre-quench Hamiltonian $H_0$, and at time $t_0 = 0$, the Hamiltonian is abruptly changed to $H$. Thereafter, the system evolves unitarily under the new Hamiltonian as
\begin{equation}
    |\Psi(t_0)\rangle = e^{-i H t_0} |\Psi_0\rangle.
\end{equation}

In CFTs, Calabrese and Cardy proposed an analytical model in which 
the initial state $|\Psi_0\rangle$ is taken to be a conformally invariant 
boundary state $|B\rangle$, which then evolves over a finite imaginary 
time $\tau'$.

\begin{equation}
    |\Psi_0\rangle \propto e^{-\tau' H} |B\rangle.
\end{equation}
Here, the parameter $\tau'$ characterizes the finite correlation length of the initial state, ensuring the regularity of the path integral in the short-distance limit. In this construction, the dynamics of the system can be studied by analytically continuing the Euclidean path integral to real time.

After obtaining the form of the initial state, we can further examine the dynamical evolution of local operators in this quenched background. In CFT, the dynamical information of the system is typically encoded in the correlation functions of local operators. In particular, for the global quench setup, the two-point correlation function
\begin{align}
    &\langle \Psi_0(t)| {\Phi}(x_0,0){\Phi}(x_1,t_1-t_0)| \Psi_0(t)\rangle\nonumber\\
    =&\langle B|e^{-\tau' H}e^{iHt_0} {\Phi}(x_0,0){\Phi}(x_1,t_1-t_0)e^{-iHt_0}e^{-\tau' H} |B\rangle\nonumber\\
    =&\langle B| {\Phi}(x_0,\tau_0){\Phi}(x_1,\tau_1)|B\rangle,
\end{align}
where the analytic continuation to real time is taken as $\tau_i \to \tau' +  i t_i$.
The above two-point correlation function is defined on a strip geometry with Euclidean width $2\tau'$, 
where both boundaries are equipped with the same conformally invariant 
boundary condition encoded by the boundary state $|B\rangle$.
This strip can be conformally mapped to the upper half-plane (UHP) 
through the transformation

\begin{equation}
    w(z) = \frac{2\tau'}{\pi}\ln z ,
\end{equation}
which allows correlation functions on the strip to be represented in terms of standard CFT correlators on the UHP.

The above correlator on the strip follows from mapping the upper half-plane (UHP) to the strip and applying Cardy's general boundary CFT expression for two-point functions of primary operators. Its extension to twist fields for computing EE was developed in the seminal work of Calabrese and Cardy \cite{Cardy:1984bb} \cite{Calabrese:2004eu}

Therefore, the two-point function on the strip can be written as 
\begin{align}
    \langle \Phi(w_0)\Phi(w_1) \rangle_{\mathrm{strip}}
    =
    \left| \frac{\partial w_0}{\partial z_0} \right|^{-2h_n}
    \left| \frac{\partial w_1}{\partial z_1} \right|^{-2h_n}
    \left(
        \frac{
            z_{0\bar{1}}\, z_{1\bar{0}}
        }{
            z_{01}\, z_{\bar{0}\bar{1}}\,
            z_{0\bar{1}}\, z_{1\bar{1}}
        }
    \right)^{2h_n}
    F(\eta),
\end{align}
where $\bar{z}_i$ denotes the complex conjugate of $z_i$, and 
$z_{i\bar{j}} = z_i - \bar{z}_j$.
The function $F(\eta)$ depends on the cross ratio
$ \eta =  \frac{z_{0\bar{0}}\, z_{1\bar{1}}}{z_{0\bar{1}}\, z_{1\bar{0}}}.$
In general, $F(\eta)$ has no closed-form expression, but its asymptotic
behavior is simple in the limits $\eta\to 0$ and $\eta\to 1$. 
When $\eta\to 1$, the bulk identity block dominates and
$F(1)=1.$ For $\eta\ll 1$, the short-distance bulk--boundary OPE yields
$ F(\eta)\simeq (A_b^{\Phi})^{2}\,\eta^{x_b},$ 
where $A_b^{\Phi}$ is the bulk-to-boundary OPE coefficient and $x_b$
is the scaling dimension of the leading boundary operator.  
For twist operators, the leading boundary operator is the identity, 
so $x_b=0$, and the expression simplifies to
$F(\eta)\simeq (A_b^{\Phi})^{2}$. For further details, see Ref.~\cite{Calabrese:2016xau}.

Now we want to consider the EE associated with the interval $A(t_0,x_0;t_1,x_1)$. We now substitute the explicit positions of the two twist operators, $w_0 = x_0 +i\tau_0$ and $w_1 = x_1 + i\tau_1$.  
In the following, after inserting the explicit coordinates and carrying out the analytic continuations $\tau_0 \rightarrow \tau' + i t_0$ and $\tau_1 \rightarrow \tau' + i t_1$, we obtain
\begin{align}
    &\langle \tilde{\sigma}_n(x_0,t_0)\, \sigma_n(x_1,t_1) \rangle_{\mathrm{strip}}\nn\\
    &=
    \left( \frac{\pi}{2\tau'} \right)^{4h_n}
    \left(
        \frac{
            \cosh{\frac{(\Delta x+\Delta t+2t_0)\pi}{4\tau'}}
            \,\cosh{\frac{(\Delta x-\Delta t-2t_0)\pi}{4\tau'}}
        }{
            4\,
            \sinh{\frac{(\Delta x-\Delta t)\pi}{4\tau'}}\,
            \sinh{\frac{(\Delta x+\Delta t)\pi}{4\tau'}}\,
            \cosh{\frac{t_0\pi}{2\tau'}}\,
            \cosh{\frac{(\Delta t+t_0)\pi}{2\tau'}}
        }
    \right)^{2h_n}
    F(\eta),
    \label{eq:correlation}
\end{align}
where $\Delta x = x_1 - x_0$ denotes the spatial separation between the two insertions, and $\Delta t = t_1 - t_0$ is their real-time separation after analytic continuation. The detailed derivation of this two-point function is presented in Appendix~\ref{sec:a}.

For $ t_0 \gg \tau'$, the cross ratio $\eta$ admits the following approximations:
 
\begin{align}
    \eta &= \frac{ 
    e^{\frac{\Delta x \pi}{2 \tau'}} \left(1 + e^{\frac{\pi t_0}{\tau'}}\right) \left(1 + e^{\frac{\pi (t_0 + \Delta t)}{\tau'}}\right)}
    {\left(e^{\frac{\Delta x \pi}{2 \tau'}} + e^{\frac{\pi (2t_0 + \Delta t)}{2 \tau'}}\right) \left(1 + e^{\frac{\pi (\Delta x + 2t_0 + \Delta t)}{2 \tau'}}\right)}\nn \\
    &\simeq\begin{cases}
        0,& \text{spacelike: }\Delta x>\Delta t,t_0<\frac{\Delta x-\Delta t}{2},\\
        1,& \text{spacelike: }\Delta x>\Delta t,t_0>\frac{\Delta x-\Delta t}{2},\\
        1,&\text{timelike: }\Delta x<\Delta t.
    \end{cases}
    \end{align}

In this paper, we analyze three types of intervals relevant to entanglement 
dynamics in CFTs: 
(i) the equal-time spacelike interval, 
(ii) the general spacelike interval with non-equal-time endpoints
(iii) the general timelike interval. 
These three configurations cover all possible spacetime conditions between the two endpoints of the subsystem.

\subsection{Equal-time spacelike case}

This case has been extensively studied in~\cite{Calabrese:2005in}. We consider the
regime $t_0\gg\tau'$ and $\Delta x-\Delta t\gg\tau'$, and focus on the equal-time
configuration by setting $t_0=t_1$. In this limit, the analytic continuation reduces to
$\tau_0=\tau_1\to\tau'+it_0$, in agreement with the prescription of
\cite{Calabrese:2007rg}. Under this continuation, the two-point function of twist
operators in~\eqref{eq:correlation} simplifies to
\begin{align}
 \langle \tilde{\sigma}_n(x_0,t_0)\, \sigma_n(x_1,t_1) \rangle_{\mathrm{strip}}
 \simeq
 \left(\frac{\pi}{2\tau'}\right)^{4h_n}
 \begin{cases}
    e^{-\frac{2h_n\pi t_0}{\tau'}},
        & t_0 < \frac{\Delta x}{2}, \\[8pt]
    e^{-\frac{h_n\pi\Delta x}{\tau'}},
        & t_0 > \frac{\Delta x}{2}.
 \end{cases}
\end{align}

The EE for the interval $A(t_0,x_0;t_0,x_1)$ then follows as
\begin{equation}
S(t_0,x_0;t_0,x_1)
\simeq
\frac{c}{3}\log\frac{2\tau'}{\pi}
+
\begin{cases}
    \frac{c\pi t_0}{6\tau'}, & t_0 < \frac{\Delta x}{2}, \\[8pt]
    \frac{c\pi\Delta x}{12\tau'}, & t_0 > \frac{\Delta x}{2},
\end{cases}
\label{eq:equal-time}
\end{equation}
reproducing the standard result for a global quench.

Equation~\eqref{eq:equal-time} exhibits two distinct dynamical regimes. At early times
($t_0<\Delta x/2$), the EE grows linearly with time, while at late
times ($t_0>\Delta x/2$) it saturates to a constant proportional to the spatial size
$\Delta x$ of the interval. These features admit a natural interpretation in terms of
the quasiparticle picture~\cite{Calabrese:2005in}.

\begin{figure}[htbp]
    \centering

    \subfigure[
 ]{
        \includegraphics[width=0.8\textwidth]{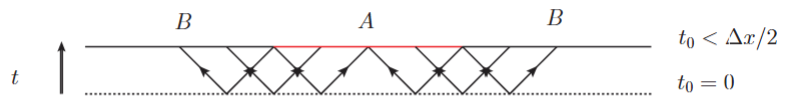}
        \label{fig:1_1}
    }
    \subfigure[ 
]{
        \includegraphics[width=0.8\textwidth]{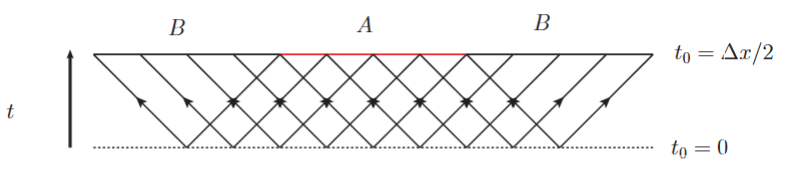}
        \label{fig:1_2}
    }

    \caption{The red line represents the interval $A$, and its complement is denoted by the interval $B$. 
    The slanted lines with arrows indicate the worldlines of an entangled pair of quasiparticles.
    (a)The evolution time has not yet reached the saturation time, $t_0 < \frac{\Delta x}{2}$.(b)The evolution time reaches the saturation time, $t_0 = \frac{\Delta x}{2}$.}
    \label{fig:3}
\end{figure}

In this picture, a global quench at $t_0=0$ produces pairs of entangled quasiparticles
that propagate in opposite directions at the speed of light. For $t_0<\Delta x/2$, one
member of each pair enters the interval $A$ while the other remains outside, leading to
a growing number of entangled pairs shared between $A$ and its complement. This process
accounts for the linear growth of entanglement, as illustrated in
Fig.~\ref{fig:3}(a).

For $t_0>\Delta x/2$, quasiparticles that entered the interval at earlier times begin to
leave, while newly produced pairs continue to arrive. These two effects balance, so
that the number of entangled pairs straddling the interval remains constant. As a
result, the EE saturates, as shown in Fig.~\ref{fig:3}(b).

\subsection{General spacelike case}
\label{section_general_spacelike_global}

Now we consider the general spacelike interval $A(t_0,x_0;t_1,x_1)$, where the two points $(t_0,x_0)$ and $(t_1,x_1)$ are spacelike separated, satisfying $\Delta x > \Delta t$. 
In the regime where $t_0 \gg \tau'$ and $\Delta x - \Delta t \gg \tau'$,the two-point correlation function \eqref{eq:correlation} can be approximated as
\begin{align}
 \langle \tilde{\sigma}_n(x_0,t_0)\, \sigma_n(x_1,t_1) \rangle_{\mathrm{strip}}
&\simeq\left(\frac{\pi}{2\tau'}\right)^{4h_n}\cdot
        \begin{cases}
            e^{-\frac{\pi h_n(\Delta t+2t_0)}{\tau'}},&t_0<\frac{\Delta x-\Delta t}{2}\\
            e^{-\frac{\pi h_n \Delta x}{\tau'}},&t_0>\frac{\Delta x-\Delta t}{2}
        \end{cases}
\end{align}
We obtain
\begin{equation}
S(t_0,x_0;t_1,x_1) \simeq  \frac{c}{3}\ln \frac{2\tau'}{\pi}  +
\begin{cases}
    \frac{c\pi(\Delta t + 2t_0)}{12\tau'},&\quad t_0 < \frac{\Delta x -\Delta t}{2} \\[8pt]
    \frac{c\pi \Delta x}{12\tau'}, &  \quad t_0 > \frac{\Delta x - \Delta t}{2}.
\end{cases}
\label{eq:general spacelike}
\end{equation}

The piecewise expression in \eqref{eq:general spacelike} shows that, after a global
quench, the EE $S(t_0,x_0;t_1,x_1)$ for a general spacelike interval
exhibits two distinct temporal regimes. At early times,
$t_0<(\Delta x-\Delta t)/2$, the EE grows linearly with time. At late
times, $t_0>(\Delta x-\Delta t)/2$, the growth terminates and
$S(t_0,x_0;t_1,x_1)$ saturates to a stationary value proportional to the spatial
separation $\Delta x$.

Compared to the equal-time configuration, the non-equal-time spacelike case displays
three essential differences. First, near $t_0=0$ the EE does not vanish
but instead takes a finite value proportional to the temporal separation $\Delta t$.
Second, the saturation time is shifted from $t_0=\Delta x/2$ to
$t_0=(\Delta x-\Delta t)/2$, indicating that saturation occurs earlier than in the
equal-time case, with the shift explicitly controlled by $\Delta t$. Third, despite the
temporal separation between the endpoints, the saturation value,
$S(t_0,x_0;t_1,x_1)=\frac{c\pi\Delta x}{12\tau'}$, depends only on the spatial separation
$\Delta x$ and is completely independent of $\Delta t$.

These features admit a natural explanation within the quasiparticle picture. A global
quench at $t_0=0$ produces pairs of entangled quasiparticles propagating in opposite
directions at the speed of light. Entanglement is generated only when exactly one
worldline of a given pair intersects the interval $A$, while its partner lies outside.
This criterion accounts for both the modified saturation time and the independence of
the late-time entanglement from the temporal separation.

\begin{figure}
    \centering
    \includegraphics[width=0.7\linewidth]{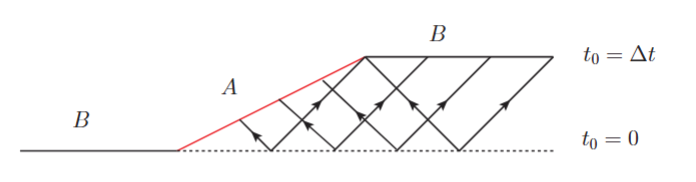}
    \caption{The red line denotes the interval $A$, while the region $B$ represents its
    entangling complement. The slanted lines with arrows indicate the worldlines of an
    entangled quasiparticle pair.}
    \label{fig:4}
\end{figure}

% First, in the short-time limit, one endpoint of the interval is located near
% $t_0 = 0$, while the other is located near $t_1 = \Delta t$. In this situation, even though the evolution time is close to zero, one of the quasiparticle worldlines already passes through the interval $A(t_0,x_0;t_1,x_1)$, while its partner worldline passes through the complement region $B$. Consequently, entanglement is already generated at this moment, and its initial value depends on the temporal separation $\Delta t$, as shown in Fig.~\ref{fig:4}.

% Second, in the equal-time case, saturation occurs when all worldlines that can cross the interval have reached their endpoints at $t_0=\Delta x/2$.  
% For the non-equal-time case with $\Delta t>0$, however, the quasiparticle worldlines intersect the entire interval at an earlier moment, causing the EE to reach saturation at $t_0 = \frac{\Delta x - \Delta t}{2}$ and therefore saturate more rapidly, with the amount of this advancement increasing with the temporal separation $\Delta t$, as shown in Fig.\ref{fig:5}(a).Third, for intervals with the same spatial length $\Delta x$, an identical number of quasiparticle worldlines cross the interval at late times, so the saturation value of the EE depends only on $\Delta x$ and is insensitive to $\Delta t$. Consequently, the EE saturates to the universal expression$S(t_0,x_0;t_1,x_1)=\frac{c\pi\,\Delta x}{12\tau'} $as illustrated in Fig.~\ref{fig:5}.

First, in the short--time limit, one endpoint of the interval lies near $t_0=0$ while the other lies near $t_1=\Delta t$. In this regime, even though the real time evolution is almost vanishing, one quasiparticle worldline already crosses the interval $A(t_0,x_0;t_1,x_1)$, while its partner lies in the complementary region $B$. As a result, entanglement is produced immediately, and its initial value depends on the temporal separation $\Delta t$, as shown in 
Fig.~\ref{fig:4}.

\begin{figure}[htbp]
    \centering

    \subfigure[]{
        \includegraphics[width=0.8\textwidth]{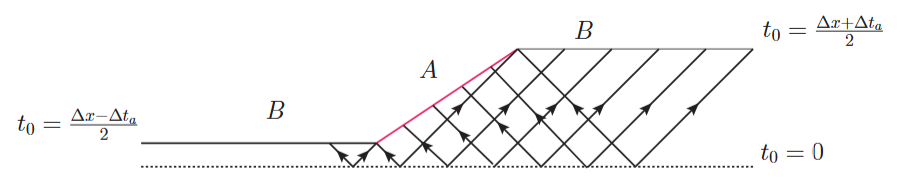}
        \label{fig:2_2}
    }

    \subfigure[]{
        \includegraphics[width=0.8\textwidth]{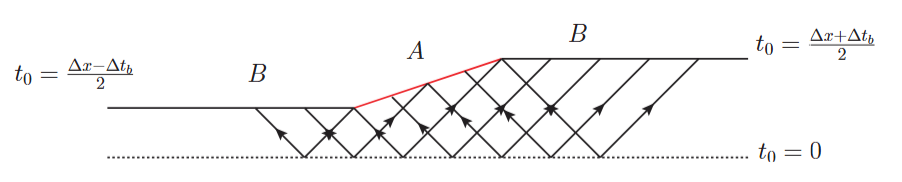}
        \label{fig:2_3}
    }

    \caption{The red line represents the interval $A$, and the region $B$ denotes its entangling partner. The slanted arrows indicate the worldlines of an entangled quasiparticle pair. We consider two temporal separations satisfying $\Delta t_a>\Delta t_b$. 
    (a) The saturation time is reached at $t_0=\frac{\Delta x+\Delta t_a}{2}$, and eight quasiparticle worldlines intersect the interval. 
    (b) The saturation time shifts to $t_0=\frac{\Delta x+\Delta t_b}{2}$, while the number of intersecting worldlines remains eight.}
    \label{fig:5}
\end{figure}

Second, in the equal--time configuration, saturation occurs when all worldlines capable of crossing the interval reach their endpoints at $t_0=\Delta x/2$. For the unequal--time configuration with $\Delta t>0$, the quasiparticle worldlines sweep across the entire interval at an earlier moment, so saturation is reached at  $ t_0=\frac{\Delta x-\Delta t}{2},$
leading to a faster approach to the plateau. The amount of advancement grows monotonically with $\Delta t$, as illustrated in Fig.~\ref{fig:5}.

Third, for intervals with the same spatial length $\Delta x$, the number of quasiparticle worldlines crossing the interval at late times is identical, so the saturation value of the EE depends only on $\Delta x$ and is insensitive to $\Delta t$. Consequently, the EE approaches the universal plateau $S(t_0,x_0;t_1,x_1)=\frac{c\pi\,\Delta x}{12\tau'},$as illustrated in Fig.~\ref{fig:5}.

\subsection{ General timelike case}

Now we consider the general timelike interval $A(t_0,x_0;t_1,x_1)$, where the two points $(t_0,x_0)$ and $(t_1,x_1)$ are timelike separated, satisfying $\Delta t > \Delta x$. 
In the regime where $t_0 \gg \tau'$ and $\Delta t - \Delta x \gg \tau'$,the two-point correlation function \eqref{eq:correlation} can be approximated as

\begin{align}
 \langle \tilde{\sigma}_n(x_0,t_0)\, \sigma_n(x_1,t_1) \rangle_{\mathrm{strip}}
&\simeq -\left(\frac{\pi}{2\tau'}\right)^{4h_n}
e^{-\frac{\pi h_n\Delta t}{\tau'}}.
\label{eq:2pt-timelike}
\end{align}

We obtain
\begin{equation}
S(t_0,x_0;t_1,x_1) \simeq  \frac{c}{3}\log \frac{2\tau'}{\pi}  +\dfrac{c\pi\Delta t }{12\tau'}+\frac{i\pi c}{6}.
\label{eq:SA-timelike}
\end{equation}

In the timelike regime, the EE exhibits three distinctive features that sharply contrast with the spacelike case.
First, the EE is already saturated at the initial time. It remains constant throughout the entire evolution, unlike the spacelike configuration, where the EE undergoes a linear growth phase before saturation.
Second, the saturated value of the EE depends only on the temporal separation $\Delta t$ and is entirely independent of the spatial distance $\Delta x$, unlike in the spacelike configuration, where the late-time EE scales linearly with the spatial distance $\Delta x$.
Finally, the EE becomes complex-valued in the timelike setup.  Moreover, the imaginary term is still the constant $i\frac{c\pi}{6}$, which is the same as the vacuum state\cite{Doi:2022iyj}.

\begin{figure}[htbp]
    \centering

    \subfigure[]{
        \includegraphics[width=0.7\textwidth]{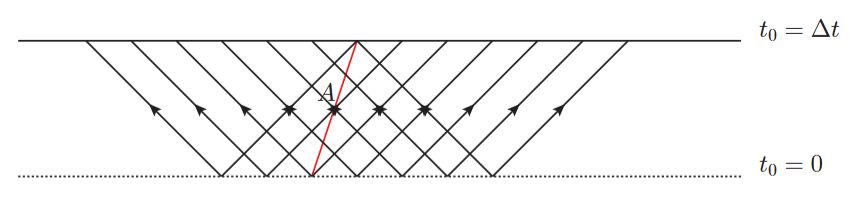}
        \label{fig:3_1}
    }

    \subfigure[]{
        \includegraphics[width=0.7\textwidth]{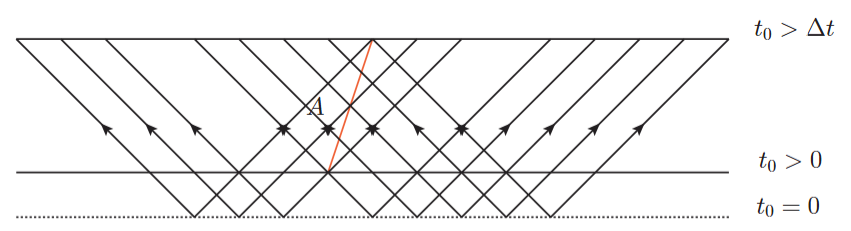}
        \label{fig:3_2}
    }

    \caption{ The red line represents the timelike interval $A$. The slanted lines with arrows indicate the worldlines of a pair of quasiparticles.(a)At the initial moment, the number of worldlines crossing the interval is eight.(b)After some evolution time, the number of quasiparticle worldlines crossing the interval remains eight.
}
    \label{fig:6}
\end{figure}

Although in the timelike configuration, a well-defined complementary region entangled with the interval $A$ can no longer be identified, the quasiparticle picture nevertheless remains a valuable tool for characterising the behaviour of the real part of the timelike EE. In this picture, a global quench at $t_0 = 0$ produces pairs of entangled quasiparticles, whose two members propagate at the speed of light in opposite directions. The entanglement dynamics is still governed by a single criterion: whether exactly one quasiparticle worldline intersects the interval $A$. In what follows, we employ this picture to account for the two characteristic features of the real part of the timelike EE discussed above.

Firstly, in a timelike configuration, the quasiparticle worldlines created by a global quench causally traverse the entire interval already at very early times. Consequently, the criterion that ``exactly one quasiparticle worldline intersects the interval'' is satisfied from the outset, and the entanglement entropy does not exhibit any linear growth regime but instead immediately saturates to its maximal value. During the subsequent evolution, the number of intersecting worldlines remains constant, so the EE is strictly time independent, as illustrated in Fig.~\ref{fig:6}.

\begin{figure}[htbp]
    \centering

    \subfigure[]{
        \includegraphics[width=0.65\textwidth]{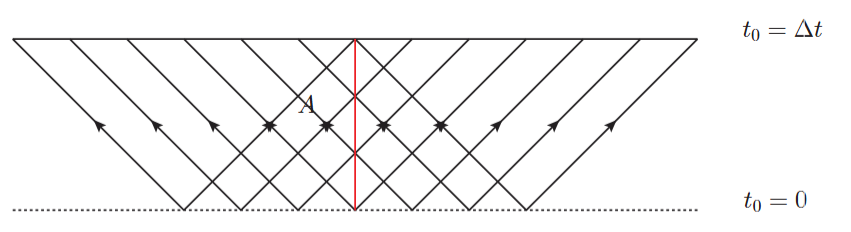}
        \label{fig:3_3}
    }

    \subfigure[]{
        \includegraphics[width=0.65\textwidth]{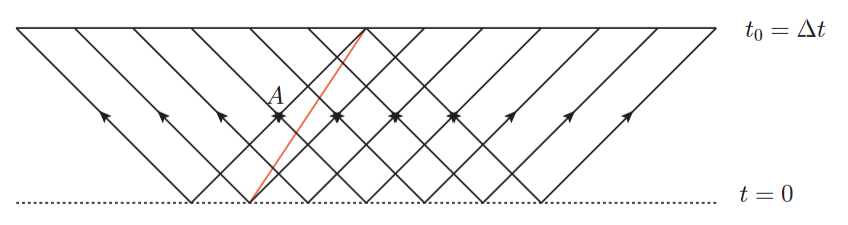}
        \label{fig:3_4}
    }

    \caption{ The red line represents the timelike interval $A$. The slanted lines with arrows indicate the worldlines of a pair of quasiparticles.(a)When $\Delta x = 0$, eight quasiparticle worldlines cross the interval.(b)When $\Delta x \neq 0$, the number of worldlines crossing the interval remains eight.}
    \label{fig:7}
\end{figure}

Secondly, for any two timelike intervals with the same temporal separation $\Delta t$, the number of quasiparticle worldlines crossing the interval is identical even if their spatial widths $\Delta x$ differ. Since the causal condition for worldline intersection depends solely on $\Delta t$, the timelike EE is a function of $\Delta t$ only and is completely insensitive to $\Delta x$, as shown in Fig.~\ref{fig:7}.

\subsection{Summary of the quasiparticle interpretation in global quench}
\label{sec:summary_quasiparticle}

The entanglement dynamics of equal-time spacelike, general spacelike, and general timelike intervals can all be accounted for by a unified quasiparticle mechanism triggered by a global quench. At $t_0=0$, the quench creates entangled pairs of excitations which propagate ballistically at the speed of light in opposite directions. Entanglement is generated whenever the worldlines of such a pair separate so that exactly one intersects the chosen interval while its partner does not. The EEis therefore proportional to the number of these straddling worldlines, and the temporal evolution of $S(t_0,x_0;t_1,x_1)$ is directly governed by the causal conditions for worldline intersection.

For an equal-time spacelike interval, quasiparticles require a finite causal time to enter the subsystem. The number of straddling trajectories therefore grows linearly until $t_0=\Delta x/2$, after which the rates of entry and exit balance, yielding saturation at a constant proportional to $\Delta x$. 
In the general spacelike configuration, a nonzero temporal separation $\Delta t$ already generates entanglement at $t_0\simeq0$, followed by a standard quasiparticle-driven evolution. The onset of saturation is shifted from $\Delta x/2$ to $(\Delta x-\Delta t)/2$. Nevertheless, the late-time EE depends only on the spatial width $\Delta x$, indicating that the spatial extent of the interval entirely controls the saturation value.
By contrast, in a timelike configuration, the quasiparticle worldlines traverse the interval essentially from the outset, and the intersection criterion is satisfied throughout the entire evolution. Consequently, no linear-growth regime develops, and the EE remains strictly time independent. In the late-time regime, the number of intersecting worldlines is controlled solely by the temporal separation $\Delta t$. It is completely insensitive to $\Delta x$, resulting in a purely temporal scaling of the real part of the timelike EE.

These findings indicate that the distinction between spacelike and timelike separations is geometric rather than dynamical: the different causal geometry alters the pattern of quasiparticle worldline intersections and thereby produces qualitatively distinct entanglement behaviour.

\section{EE in local quench model}
\label{sec:4}
A local quench describes a nonequilibrium evolution triggered by a localized perturbation acting on a finite spatial region. Such an excitation can be realized by inserting a local operator on the vacuum at position $x$ and time $t=0$, producing a locally excited state,
\begin{equation}
    |\Psi\rangle \propto \mathcal{O}(x,0)\,|0\rangle ,
\end{equation}
where $\mathcal{O}(x,0)$ is a local operator inserted at the spacetime point $(x,0)$.
The resulting state subsequently evolves under the Hamiltonian.
The corresponding density matrix of the locally excited state takes the form
\begin{align}
    \rho^{\mathrm{EX}}
    &= \mathcal{N}\,  e^{-\epsilon H}\, O(x,0)\, |0\rangle \langle 0|\,
       O(x,0)\, e^{-\epsilon H} , 
\end{align}
where $\mathcal{N}$ ensures $\mathrm{Tr}\,\rho^{\mathrm{EX}}=1$. Here, the small parameter $\epsilon > 0$ acts as a UV regulator, which slightly separates the operator insertions in imaginary time to ensure convergence and regularize short-distance singularities. Finally, we take the limit $\epsilon\to 0$. Here, we will consider the EE associated with the interval $A(t_0,0;t_1, L)$ between $(t_0,0)$ and $(t_1, L)$.

We begin by performing a Wick rotation to Euclidean time, which renders the path-integral representation of the density matrix well defined. Once the theory is formulated in Euclidean space, the Rényi entropies can be computed using the replica method: $\mathrm{Tr}\,\rho_A^n$ is represented as a correlation function of twist operators on an $n$-sheeted Euclidean manifold.
For the interval whose Lorentzian endpoints are $(t_0,0)$ and $(t_1,L)$, the Wick rotation maps it to the Euclidean interval
$ A_E(\tau_0,0;\,\tau_1,L),$where $\tau_0$ and $\tau_1$ denote the Euclidean times. We then carry out the replica computation entirely in Euclidean signature. The  excess Rényi entropy is given by
\begin{align}\label{DeltaSn}
    \Delta S^{(n)}(\tau_0,0;\,\tau_1,L)
    &= S^{(n)}\!\left(\rho^{\mathrm{EX}}(\tau_0,0;\,\tau_1,L)\right)
       - S^{(n)}\!\left(\rho^{(0)}(\tau_0,0;\,\tau_1,L)\right) \notag\\
    &= \frac{1}{1 - n}
    \log
    \frac{
        \big\langle
        O(w_1,\bar{w}_1)\,
        O(w_2,\bar{w}_2)\cdots
        O(w_{2n},\bar{w}_{2n})
        \big\rangle_{\Sigma_n}
    }{
        \left(
        \big\langle
        O(w_1,\bar{w}_1)\,
        O(w_2,\bar{w}_2)
        \big\rangle_{\Sigma_1}
        \right)^n
    } .
\end{align}
Here the two operators $O(w_1,\bar w_1)$ and $O(w_2,\bar w_2)$ are inserted at the same
spatial position $x$, with complex coordinates
\begin{equation}
     w_1 = x + i\epsilon, \qquad \bar w_1 = x - i\epsilon, \qquad
    w_2 = x - i\epsilon, \qquad \bar w_2 = x + i\epsilon .
\end{equation}
For the locally excited state, these two insertion points are replicated on each
sheet of the $n$-sheeted replica manifold $\Sigma_n$, thereby generating the set$\{w_1,\ldots,w_{2n}\}$ that appears in the corresponding $2n$-point correlation function.The detailed derivation of $\Delta S^{(n)}(\tau_0,0;\,\tau_1,L)$ is presented in Appendix~\ref{sec:b}. Finally, the Lorentzian EE is obtained by analytically continuing the Euclidean result according to $\tau_0 \to i t_0, \, \tau_1 \to i t_1 .$

Let us first consider the second R\'enyi entropy $\Delta S^{(2)}(t_0,x_0;t_1,x_1)$. Here we focus on rational CFTs, since the result for the second R\'enyi entropy can be directly generalized to the $n$-th R\'enyi entropy. 
This follows from the quasiparticle nature of the local excitation in rational CFTs.

We consider an interval $A_E(\tau_0,0;\tau_1,L)$, for which the correlation functions
appearing in the numerator of Eq.~(\ref{DeltaSn}) are defined on a two-sheeted
Riemann surface $\Sigma_2$ branched at the endpoints of the interval.
This surface can be uniformized to a single complex plane $\Sigma_1$
with coordinates $(z,\bar z)$ by the conformal transformation
\begin{equation}
    z_i = \sqrt{\frac{w_i - w_L}{\,w_i - w_R\,}}, \qquad
    \bar z_i = \sqrt{\frac{\bar w_i - \bar w_L}{\,\bar w_i - \bar w_R\,}},
\end{equation}
where the endpoints of the interval in the $w$-coordinate are given by
$w_L = i\tau_0$ and $w_R = L + i\tau_1$. After performing the analytic continuation
$\tau_0 \to i t_0$ and $\tau_1 \to i t_1$, the insertion points are mapped to

\begin{equation}
    z_1 = -z_3
    = \sqrt{
        \frac{x + t_0 + i\epsilon}
             {x - L + t_1 + i\epsilon}
    }, \qquad
    \bar z_1 = -\bar z_3
    = \sqrt{
        \frac{x - t_0 - i\epsilon}
             {x - L - t_1 - i\epsilon}
    },
\end{equation}
and
\begin{equation}
    z_2 = -z_4
    = \sqrt{
        \frac{x + t_0 - i\epsilon}
             {x - L + t_1 - i\epsilon}
    }, \qquad
    \bar z_2 = -\bar z_4
    = \sqrt{
        \frac{x - t_0 + i\epsilon}
             {x - L - t_1 + i\epsilon}
    } .
\end{equation}

This conformal mapping removes the branch cut between the interval endpoints
and reduces the computation of the second Rényi entropy to the evaluation of
correlation functions on a single complex plane.

Moreover, if we further set $t_0 = t_1$, the above transformation reduces precisely to the original conformal map given in \cite{He:2014mwa}. 
This serves as a consistency check, confirming the validity of the set-up in this paper.

After performing the conformal mapping introduced above, the correlator takes the form:

\begin{align}
    &\frac{\langle O(w_1,\bar{w}_1)O(w_2,\bar{w}_2)
    O(w_3,\bar{w}_3)O(w_4,\bar{w}_4)\rangle_{\Sigma_2}}
    {\left(\langle O(w_1,\bar{w}_1)O(w_2,\bar{w}_2)\rangle_{\Sigma_1}\right)^2}
    =|\eta|^{4h}|1-\eta|^{4h}\cdot G(\eta,\bar{\eta}),
    \label{eq:4pt-Sigma2}
\end{align}
where $G(\eta,\bar{\eta})=\displaystyle 
\langle \mathcal O(0)\,\mathcal O(\eta)\,\mathcal O(1)\,\mathcal O(\infty)\rangle $, $\eta$ and $\bar \eta$ are the cross ratios:
\begin{align}
    \eta = \frac{z_{12}z_{34}}{z_{13}z_{24}}, \quad \bar{\eta} = \frac{\bar{z}_{12}\bar{z}_{34}}{\bar{z}_{13}\bar{z}_{24}}.
\end{align}
Finally, by analytically continuing the Euclidean insertion points
$\tau_0 \rightarrow   i t_0, \,\tau_1 \rightarrow   i t_1,$we obtain the Lorentzian-time expressions for the cross ratios $\eta$ and
$\bar{\eta}$. Substituting these into the four-point function above, then directly gives the value of the second R\'enyi entropy $\Delta S^{(2)}(t_0,x_0;t_1,x_1)$.

\subsection{Example: massless free scalar field}

To proceed, we consider a case of a $c=1$ CFT 
described by a massless free scalar field $\phi$. 
The local operator is chosen as
\begin{align}
    O = \frac{1}{\sqrt{2}}\left(e^{\frac{i}{2}\phi} + e^{-\frac{i}{2}\phi}\right),
\end{align}
which has conformal dimensions $h = \bar{h} = 1/8$.

\vspace{0.5em}

For this model, the four-point function takes the explicit form
\begin{equation}
    G(\eta,\bar{\eta}) = \frac{|\eta| + 1 + |1 - \eta|}{2\sqrt{|\eta||1 - \eta|}} .
\end{equation}
Substituting this expression into (\ref{DeltaSn}) and (\ref{eq:4pt-Sigma2}) yields
\begin{eqnarray}
    \Delta S^{(2)}(0,0;\Delta t,L) 
    =-\log\frac{|\eta|+1+|1-\eta|}{2}
\end{eqnarray}
In the limit $\epsilon \to 0$, we find
\begin{align}
    \eta&=
    \begin{cases}
        0, & t_0\in[0\,,-x) \cup (-x + L - \Delta t\,,+\infty), \\
        1, & t_0\in(-x\,,-x + L - \Delta t).
    \end{cases}
\end{align}

\begin{align}
    \bar{\eta}
    &=\begin{cases}
        0, &t_0\in[0\,,x-L-\Delta t) \cup (x\,,+\infty),\\ 
        1, &t_0\in(x-L-\Delta\,,x).
    \end{cases}
\end{align}
The detailed derivation of this result is provided in Appendix~\ref{sec:c}.

As a result, the second R\'enyi entropy is given by

\begin{align}
     \Delta S^{(2)}(0,0;\Delta t,L)
     =\begin{cases}
        \log 2, & (\eta,\bar{\eta}) = (0,1) \text{ or } (1,0), \\
        0, & (\eta,\bar{\eta}) = (0,0) \text{ or } (1,1).
    \end{cases}
\end{align}

This shows that the entanglement increases by $\log 2$ precisely when the time evolution drives the cross ratios to $(\eta,\bar{\eta})=(0,1)$ or $(1,0)$, while remaining zero in the other two cases; the physical origin of these conditions will be clarified using the quasiparticle picture below.  Note that the above results are applied to a general interval $A(0,0;\Delta t, L)$, which can be both spacelike or timelike.

\subsection{General primary operators in rational CFTs}

In rational CFTs,  for the general primary operator $O_a$,
one finds the following universal relations:
\begin{align}
    &\frac{\langle O_a(w_1,\bar{w}_1)O_a(w_2,\bar{w}_2)
    O_a(w_3,\bar{w}_3)O_a(w_4,\bar{w}_4)\rangle_{\Sigma_2}}
    {\left(\langle O_a(w_1,\bar{w}_1)O_a(w_2,\bar{w}_2)\rangle_{\Sigma_1}\right)^2}
    \notag\\
    &=\begin{cases}
        1/d_a, & (\eta,\bar{\eta}) = (0,1)\ \text{or}\ (1,0), \\[4pt]
        1, & (\eta,\bar{\eta}) = (0,0)\ \text{or}\ (1,1),
    \end{cases}    
\end{align}
where $d_a$ denotes the quantum dimension of the operator $O_a$, 
which is related to the modular $S$-matrix by
$d_a = \frac{S_{0a}}{S_{00}}$ .
The results can be found in \cite{He:2014mwa}

Correspondingly, the excess of the second R\'enyi entropy can be written as
\begin{align}
     \Delta S^{(2)}(0,0;\Delta t,L)
     &=\begin{cases}
        \log d_a, & (\eta,\bar{\eta}) = (0,1)\ \text{or}\ (1,0), \\[4pt]
        0, & (\eta,\bar{\eta}) = (0,0)\ \text{or}\ (1,1).
    \end{cases}
\end{align}

This indicates the second R\'enyi entropy is universally determined by the quantum dimension of the primary operator.

Likewise, the $n$-th R\'enyi entropy $S^{(n)}(0,0;\Delta t, L)$ displays the same universal structure, with its excess fixed entirely by the quantum dimension $d_a$.

\subsection{Equal-time spacelike interval}

In the equal-time spacelike configuration, the insertion of a local operator can be categorized into four distinct cases depending on its position relative to the entangling interval.

\textbf{Case 1:} $x <0$ —  The operator is inserted at a point with  $x <0$. 
Causality between the insertion point and region $A(0,0;\Delta t,L)$ is established for 
$t_0 \in (-x,\, -x + L)$, leading to a temporary increase in the EE by $\log d_a$.
\begin{align}
    \Delta S^{(n)}(0,0;\Delta t,L) =
    \begin{cases}
        0, & t_0 \in [0, -x) \cup (-x + L, +\infty), \\[4pt]
        \log d_a, & t_0 \in (-x,\, -x + L).
    \end{cases}
\end{align}

\textbf{Case 2:} $0 < x < \frac{L}{2}$ — The operator is inserted at a point with $0 < x < \frac{L}{2}$. Causality between the insertion and the exterior of $A(0,0;\Delta t,L)$ is established for 
$t_0 \in (x,\, -x + L)$, leading to a temporary increase in the EE by $\log d_a$.
\begin{align}
    \Delta S^{(n)}(0,0;\Delta t,L) =
    \begin{cases}
        0, & t_0 \in [0, x) \cup (-x + L, +\infty), \\[4pt]
        \log d_a, & t_0 \in (x,\, -x + L).
    \end{cases}
\end{align}

\textbf{Case 3:} $\frac{L}{2} < x < L$ — The operator is inserted at a point with $\frac{L}{2} < x < L$.
Causality between the insertion and the opposite boundary of $A(0,0;\Delta t,L)$ is established for 
$t_0 \in (-x + L,\, x)$, leading to a temporary increase in the EE by $\log d_a$.
\begin{align}
    \Delta S^{(n)}(0,0;\Delta t,L) =
    \begin{cases}
        0, & t_0 \in [0, -x + L) \cup (x, +\infty), \\[4pt]
        \log d_a, & t_0 \in (-x + L,\, x).
    \end{cases}
\end{align}

\textbf{Case 4:} $x > L$ — The operator is inserted at a point with $x > L$.
Causality between the insertion and region $A(0,0;\Delta t,L)$ is established for 
$t_0 \in (x - L,\, x)$, leading to a temporary increase in the EE by $\log d_a$.
\begin{align}
    \Delta S^{(n)}(0,0;\Delta t,L) =
    \begin{cases}
        0, & t_0 \in [0, x - L) \cup (x, +\infty), \\[4pt]
        \log d_a, & t_0 \in (x - L,\, x).
    \end{cases}
\end{align}

\begin{figure}
    \centering
    \includegraphics[width=0.75\linewidth]{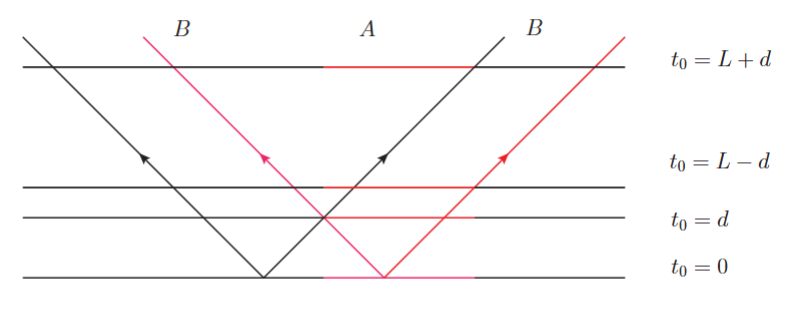}
    \caption{The red horizontal segment represents the interval $A$, whose complement is denoted by $B$, with endpoints at $(0,0;0,L)$ at the initial time. The black slanted lines correspond to the quasiparticle worldlines generated by an insertion at $x=-d$ and realize Type~I, whereas the red slanted lines correspond to the worldlines generated by an insertion at $x=d$ and realize Type~II. Here $d<L$.}
    \label{fig:8}
\end{figure}

We now account for this result within the quasiparticle picture. The operator insertion at $t=0$ is treated as creating an entangled quasiparticle pair propagating in opposite directions at the speed of light. Whenever, at a given time, the two worldlines are arranged such that one intersects the interval $A$ while the other lies in its complement $B$, the entanglement receives a positive contribution. If both worldlines intersect $A$, or if neither intersects $A$, no entanglement is generated. These four cases naturally fall into two Types; see Fig.~\ref{fig:8}.

Cases~1 and~4 belong to Type~I, in which no quasiparticle worldline intersects the interval $A$ at the initial time. The black segment indicates the allowed insertion region. In this regime no entanglement is generated initially. As the evolution proceeds, exactly one worldline eventually crosses the interval, thereby generating entanglement and contributing $\log d_a$. At later times, the interval separates from the worldline again, and this contribution disappears.

Cases~2 and~3 belong to Type~II, in which both quasiparticle worldlines intersect the interval $A$ at the initial time. The red segment indicates the allowed insertion region. In this regime no entanglement is generated initially. As the evolution proceeds, one worldline eventually exits the interval while the other still intersects it, thereby producing entanglement and contributing $\log d_a$. At later times, as the evolution continues, both worldlines move away from the interval, and this contribution disappears.

\subsection{General spacelike interval}\label{Section_general_spacelike_local}
In the general spacelike configuration, the insertion of a local operator can be categorized into five distinct cases depending on its position relative to the entangling interval. 

\textbf{Case 1:} $x < 0$ — The operator is inserted at a point with $x < 0$. 
Causality between the insertion point and region $A(0,0;\Delta t,L)$ is established for 
$t_0 \in (-x,\, -x + L - \Delta t)$, leading to a temporary increase in the EE by $\log d_a$.
\begin{align}
    \Delta S^{(n)}(0,0;\Delta t,L) = 
    \begin{cases}
        0, & t_0 \in [0, -x) \cup (-x + L - \Delta t, +\infty), \\[4pt]
        \log d_a, & t_0 \in (-x,\, -x + L - \Delta t).
    \end{cases}
\end{align}

\textbf{Case 2:} $0 < x < \frac{L - \Delta t}{2}$ —  The operator is inserted at a point with  $0 < x < \frac{L - \Delta t}{2}$.
Causality between the insertion and the exterior of $A(0,0;\Delta t,L)$ is established for 
$t_0 \in (x,\, -x + L - \Delta t)$, leading to a temporary increase in the EE by $\log d_a$.
\begin{align}
    \Delta S^{(n)}(0,0;\Delta t,L) = 
    \begin{cases}
        0, & t_0 \in [0, x) \cup (-x + L - \Delta t, +\infty), \\[4pt]
        \log d_a, & t_0 \in (x,\, -x + L - \Delta t).
    \end{cases}
\end{align}

\textbf{Case 3:} $\frac{L - \Delta t}{2} < x < L - \Delta t$ —The operator is inserted at a point with  $\frac{L - \Delta t}{2} < x < L - \Delta t$ .
Causality between the insertion and the opposite side of $A(0,0;\Delta t,L)$ is established for 
$t_0 \in (-x + L - \Delta t,\, x)$, leading to a temporary increase in the EE by $\log d_a$.
\begin{align}
    \Delta S^{(n)}(0,0;\Delta t,L) = 
    \begin{cases}
        0, & t_0 \in [0, -x + L - \Delta t) \cup (x, +\infty), \\[4pt]
        \log d_a, & v \in (-x + L - \Delta t,\, x).
    \end{cases}
\end{align}

\textbf{Case 4:} $L - \Delta t < x < L + \Delta t$ — The operator is inserted at a point with $L - \Delta t < x < L + \Delta t$.
Causality between the insertion and region $A(0,0;\Delta t,L)$ is established for 
$t_0 \in [0,\, x)$, leading to a temporary increase in the EE by $\log d_a$.
\begin{align}
    \Delta S^{(n)}(0,0;\Delta t,L) = 
    \begin{cases}
        0, & t_0 \in (x, +\infty), \\[4pt]
        \log d_a, & t_0 \in [0,\, x).
    \end{cases}
\end{align}

\textbf{Case 5:} $x > L + \Delta t$ — The operator is inserted at a point with $x > L + \Delta t$.
Causality between the insertion point and region $A(0,0;\Delta t,L)$ is established for 
$t_0 \in (x - L - \Delta t,\, x)$, leading to a temporary increase in the EE by $\log d_a$.
\begin{align}
    \Delta S^{(n)}(0,0;\Delta t,L) = 
    \begin{cases}
        0, & t_0 \in [0, x - L - \Delta t) \cup (x, +\infty), \\[4pt]
        \log d_a, & t_0 \in (x - L - \Delta t,\, x).
    \end{cases}
\end{align}

\begin{figure}
    \centering
    \includegraphics[width=0.85\linewidth]{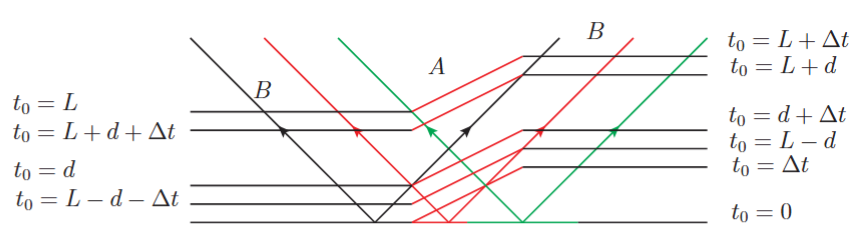}
    \caption{The red horizontal segment represents the interval $A$, whose complement is denoted by $B$, with endpoints at $(0,0;\Delta t,L)$ at the initial time. The black slanted lines correspond to the quasiparticle worldlines generated by an insertion at $x=-d$ and realize Type~I, whereas the red slanted lines correspond to the worldlines generated by an insertion at $x=d$ and realize Type~II. The green slanted lines correspond to the worldlines generated by an insertion at $x=L$ and realize Type~III. Here $\Delta t<L$ and $0<d<L$.}
    \label{fig:9}
\end{figure}

We now account for this result within the quasiparticle picture. The operator insertion at $t=0$ is treated as creating an entangled quasiparticle pair propagating in opposite directions at the speed of light. Whenever, at a given time, the two worldlines are arranged such that one intersects the interval $A$ while the other lies in its complement $B$, the entanglement receives a positive contribution. If both worldlines intersect $A$, or if neither intersects $A$, no entanglement is generated. These five cases naturally fall into three Types; see Fig.~\ref{fig:9}.

Cases~1 and~5 belong to Type~I, in which no quasiparticle worldline intersects the interval $A$ at the initial time. The black segment indicates the allowed insertion region. In this regime, no entanglement is generated initially. As the evolution proceeds, exactly one worldline eventually crosses the interval, thereby generating entanglement and contributing $\log d_a$. At later times, the interval separates from the worldline again, and this contribution disappears.

Cases~2 and~3 belong to Type~II, in which both quasiparticle worldlines intersect the interval $A$ at the initial time. The red segment indicates the allowed insertion region. In this regime no entanglement is generated initially. As the evolution proceeds, one worldline eventually exits the interval while the other still intersects it, thereby producing entanglement and contributing $\log d_a$. At later times, as the evolution continues, both worldlines move away from the interval, and this contribution disappears.

Case~4 belongs to Type~III, in which one quasiparticle worldline intersects the interval $A$ at the initial time while the other does not. The green segment indicates the allowed insertion region. In this regime entanglement is generated already at $t_0=0$, contributing $\log d_a$. As the evolution proceeds, the intersecting worldline eventually exits the interval, so that both worldlines move away from $A$, and this contribution disappears.

\subsection{General timelike interval}\label{Section_general_timelike_local}

In the general timelike configuration, the insertion of a local operator can be categorized into four distinct cases depending on its position relative to the entangling interval.

\textbf{Case 1:} $x < L - \Delta t$ — The operator is inserted at a point with $x < L - \Delta t$.
Causality between the excitation and region $A(0,0;\Delta t,L)$ is established for 
$t_0 \in (-x + L - \Delta t,\, -x)$, during which the EE temporarily increases by $\log d_a$.
\begin{align}
    \Delta S^{(n)}(0,0;\Delta t,L) = 
    \begin{cases}
        0, & t_0 \in [0,\, -x + L - \Delta t) \cup (-x,\, +\infty), \\[4pt]
        \log d_a, & t_0 \in (-x + L - \Delta t,\, -x).
    \end{cases}
\end{align}

\textbf{Case 2:} $L - \Delta t < x < 0$ — The operator is inserted at a point with $L - \Delta t < x < 0$.
Causality between the insertion and region $A(0,0;\Delta t,L)$ is established for 
$t_0 \in [0,\, -x)$, leading to a temporary increase in the EE by $\log d_a$.
\begin{align}
    \Delta S^{(n)}(0,0;\Delta t,L) = 
    \begin{cases}
        0, & t_0 \in (-x,\, +\infty), \\[4pt]
        \log d_a, & t_0 \in [0,\, -x).
    \end{cases}
\end{align}

\textbf{Case 3:} $0 < x < L + \Delta t$ — The operator is inserted at a point with $0 < x < L + \Delta t$.
Causality between the insertion and region $A(0,0;\Delta t,L)$ is established for 
$t_0 \in [0,\, x)$, leading to a temporary increase in the EE by $\log d_a$.
\begin{align}
    \Delta S^{(n)}(0,0;\Delta t,L) = 
    \begin{cases}
        0, & t_0 \in (x,\, +\infty), \\[4pt]
        \log d_a, & t_0 \in [0,\, x).
    \end{cases}
\end{align}

\textbf{Case 4:} $x > L + \Delta t$ — The operator is inserted at a point with $x > L + \Delta t$.
Causality between the insertion and region $A(0,0;\Delta t,L)$ is established for 
$t_0 \in (x - L - \Delta t,\, x)$, leading to a temporary increase in the EE by $\log d_a$.

\begin{align}
    \Delta S^{(n)}(0,0;\Delta t,L) = 
    \begin{cases}
        0, & t_0 \in [0,\, x - L - \Delta t) \cup (x,\, +\infty), \\[4pt]
        \log d_a, & t_0 \in (x - L - \Delta t,\, x).
    \end{cases}
\end{align}

\begin{figure}
    \centering
    \includegraphics[width=0.8\linewidth]{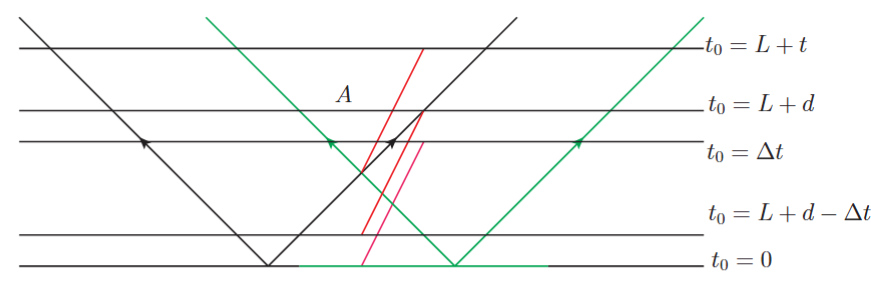}
    \caption{The red horizontal segment represents the interval $A$, whose complement is denoted by $B$, with endpoints at $(0,0;\Delta t,L)$ at the initial time. The black slanted lines correspond to the quasiparticle worldlines generated by an insertion at $x=-d$ and realize Type~I, whereas the green slanted lines correspond to the worldlines generated by an insertion at $x=d$ and realize Type~III. Here $\Delta t>L$ and $0<d<L$.}
    \label{fig:10}
\end{figure}

We now account for this result within the quasiparticle picture. The operator insertion at $t=0$ is treated as creating an entangled quasiparticle pair propagating in opposite directions at the speed of light. Whenever, at a given time, the two worldlines are arranged such that one intersects the interval $A$ while the other lies outside $A$, the entanglement receives a positive contribution. If both worldlines intersect $A$, or if neither intersects $A$, no entanglement is generated. These four cases naturally fall into two Types; see Fig.~\ref{fig:10}.

Cases~1 and~4 belong to Type~I, in which no quasiparticle worldline intersects the interval $A$ at the initial time. The black segment indicates the allowed insertion region. In this regime, no entanglement is generated initially. As the evolution proceeds, exactly one worldline eventually crosses the interval, thereby generating entanglement and contributing $\log d_a$. At later times the interval separates from the worldline again, and this contribution disappears.

Cases~2 and~3 belong to Type~III, in which one quasiparticle worldline intersects the interval $A$ at the initial time while the other does not. The green segment indicates the allowed insertion region. In this regime, entanglement is generated already at $t_0=0$, contributing $\log d_a$. As the evolution proceeds, the intersecting worldline eventually exits the interval, so that both worldlines move away from $A$, and this contribution disappears.

\subsection{Summary of the quasiparticle interpretation in local quench}
\label{Section_quasiparticle_global}

%In this subsection, we demonstrate that the evolution of EE generated by a local operator insertion can be consistently interpreted using the quasiparticle picture, regardless of whether the interval is equal-time spacelike, general spacelike, or general timelike. In all cases, the key principle remains the same: the time dependence of the entanglement is determined solely by whether and when exactly one worldline of each quasiparticle pair crosses the interval. At $t_0=0$, a local operator insertion creates a single pair of quasiparticles at the excitation point, which then propagate in opposite directions at the speed of light. However, the conditions under which the interval intersects the quasiparticle worldlines differ from case to case, thereby giving rise to distinct dynamical behaviors.

The entanglement dynamics produced by a local operator insertion admit a unified quasiparticle description for equal-time spacelike intervals, general spacelike 
intervals, and general timelike intervals. In all configurations, an insertion at 
$t_0=0$ excites a pair of entangled quasiparticles that propagate at the speed of 
light. Depending on the position of the insertion, the causal relation between the quasiparticle worldlines and the interval $A$ at the initial time falls into three Types: (I) both worldlines intersect $A$; (II) neither worldline intersects $A$; (III) exactly one worldline intersects $A$ while the other does not.

%For equal--time spacelike intervals, the insertion point only realizes Type~1 and Type~2. For general spacelike intervals, the causal domains of the two endpoints are asymmetric, and Type~3 emerges in addition to Type~1 and Type~2. For general timelike intervals, the endpoints are timelike--separated, so only Type~1 and Type~3 are realized, whereas Type~2 is absent.

The criterion for entanglement production is completely universal across the three interval geometries: entanglement increases only when one quasiparticle worldline intersects the interval while the other does not; if both worldlines 
lie inside or outside the interval, no entanglement is produced.

\section{Relation between time- and spacelike EE}\label{sec:5}

In the previous section, we obtained the exact EE for general intervals $A(t_0,x_0;t_1,x_1)$ in dynamical states, demonstrating that the timelike EE is also well-defined and computable in generic situations. Earlier studies of time- and spacelike EE in vacuum and thermal states suggest that certain relations exist between the two \cite{Guo:2024lrr}. In this section, we aim to explore whether similar relations hold in dynamical states. Indeed, we will show that such relations also arise in the two types of states considered here.

\subsection{Review of the relation for the vacuum state}

As we have shown in Sec.~\ref{sec:3} and Sec.~\ref{sec:4}, the timelike EE can be obtained from correlators of twist operators with timelike separation, namely \eqref{eq:correlation}. In QFTs, causality constraints require that timelike correlators be related to spacelike correlators supported in suitable subregions on a given Cauchy surface. This can also be understood from the viewpoint of algebraic QFTs: for a given subregion $A$ on a Cauchy surface, the algebra $\mathcal{R}(A)$ is equal to the algebra of its causal domain $\mathcal{D}(A)$. However, in general it is not possible to write down an explicit relation that expresses operators in $\mathcal{D}(A)$ solely in terms of operators on $A$\footnote{For example, in free QFTs with field $\phi$, one may use canonical quantization to decompose the operator $\phi(t,\vec{x})$ in terms of $\phi(0,\vec{y})$ and its conjugate momentum $\pi(0,\vec{y})$ on the Cauchy surface $t=0$.}.

Here we focus on timelike and spacelike EE, which are determined by twist operators and their two-point correlation functions. In 2D CFTs, the authors of \cite{Guo:2024lrr}
showed that an exact relation between time- and spacelike EE can be constructed for the vacuum state.  Consider the general interval $A(t,x;t',x')$ with the ending points $(t,x)$ and $(t',x')$.  The past lightcone of these two points would have four intersections between the timeslice $t=t_0$ (assumed $t_0<t'<t$) as shown in Fig.\ref{fig:light_cones}. In the vacuum state, we can build the following relation:

\begin{figure}
    \centering
    \includegraphics[width=0.5\linewidth]{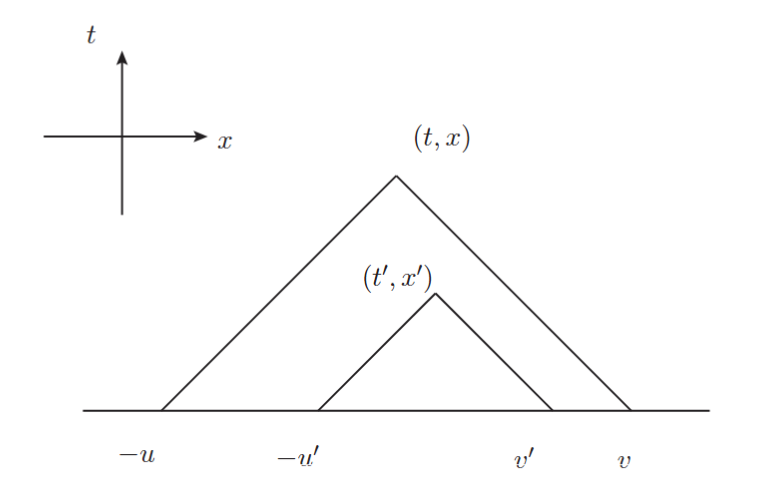}
    \caption{In a typical scenario where $(t,x)$ and $(t',x')$ are timelike, their past light cones intersect with four points at $t=t_0$ with $-u < -u' < v' < v$.}
    \label{fig:light_cones}
\end{figure}

\begin{align}
S(t,x;t',x') 
&= \tfrac{1}{4}\Big[ S(t_0,-u;\,t_0,-u') + S(t_0,-u;\,t_0,v') + S(t_0,v;\,t_0,-u') + S(t_0,v;\,t_0,v') \Big] \nonumber\\ 
&\quad + \tfrac{1}{4}\!\int_{-u'}^{v'} \!dy\, \partial_{t'} S(t_0,u;\,t_0,y)
+ \tfrac{1}{4}\!\int_{-u'}^{v'} \!dy\, \partial_{t'} S(t_0,v;\,t_0,y) \nonumber\\
&\quad + \tfrac{1}{4}\!\int_{-u}^{v} \!dy\, \partial_{t} S(t_0,y;\,t_0,-u')
+ \tfrac{1}{4}\!\int_{-u}^{v} \!dy\, \partial_{t} S(t_0,y;\,t_0,v') \nonumber\\
&\quad + \tfrac{1}{4}\!\int_{-v}^{v} \!dy \int_{-v'}^{v'} \!dy' \,\partial_t\partial_{t'} S(t_0,y;\,t_0,y'),
\label{eq:vacuum-decomposition}
\end{align}
where $u=(t-t_0)-x$, $v=(t-t_0)+x$, $u'=(t'-t_0)-x'$ and $v'=(t'-t_0)+x'$. In the above expressions, the right-hand side contains only the EE of spacelike subregions on the Cauchy surface $t = t_0$ within the interval $[-u, v]$, together with their temporal derivatives. This result indicates that the EE for timelike intervals can be reconstructed from spacelike entanglement data on a fixed Cauchy slice. In the vacuum state, one can prove this relation using the two-point functions of a 2D free scalar theory. In 2D CFTs, one can further check that the relation still holds for a large class of states that admit a gravity dual in pure AdS$_3$. For general states, one expects this relation to be modified. However, as we will show below, the same relation continues to hold for the dynamical states considered in this work, thereby providing a nontrivial extension of \eqref{eq:vacuum-decomposition} to more general cases. 

In the following, we will consider a specific case, that is $t'=t_0$, $t=t_1$ and $x=x'=0$, that is an interval $(t_0,t_1)$ on the time coordinate, i.e., $A(t_0,0;t_1,0)$. In this case, we have $u=v=t_1-t_0:=\Delta t$ and $u'=v'=0$. In this case, the relation would be much simpler, 
\begin{eqnarray}\label{relation_simpler}
    S(t_0,0;t_1,0)= \frac{1}{2} \left[S(t_0,-\Delta t;\,t_0,0) + S(t_0,\Delta t;\,t_0,0)  \right]+\frac{1}{2}\int_{-\Delta t}^{\Delta t}dy \partial_t S(t_0,y;t_0,0),
\end{eqnarray}
where the term $\partial_t S(t_0,y;t_0,0):=\partial_t S(t,y;t_0,0)|_{t\to t_0}$.In the previous section, we have already computed $S(t,x;t',x')$ for an arbitrary interval $A(t,x;t',x')$. We will directly verify whether the above relation applies to the states in these examples. 

\subsection{Relation for global quench state}

For the timelike interval $A(t_0,0;t_1,0)$, the EE is given by 
\bea
S(t_0,0;t_1,0)=\frac{c}{3}\log \frac{2\tau'}{\pi}  +\dfrac{c\pi\Delta t }{12\tau'}+\frac{i\pi c}{6}.
\eea
The terms on the right-hand side of (\ref{relation_simpler}) are only related to the spacelike intervals. In the vacuum state, each term would be constant; the temporal derivative term contributes only the imaginary part of the timelike EE. However, for the dynamical state we consider here, each term on the right-hand side would be time-dependent. Thus, this would be a highly non-trivial extension of the relation to dynamical states.

Let us first consider the terms with a temporal derivative. In the dynamical state, the EE of an interval would be time-dependent; thus, the temporal derivative of EE is generally non-vanishing, which can be understood as the ``velocity'' of the EE evolution.  For an interval at timeslice $t=t_0$ with length $\Delta x$, by using \eqref{eq:general spacelike} we have
\begin{align}
\partial_{t} S(t_0,\Delta x;t_0,0) = &\frac{c\pi}{24\tau'} \!\big[
    \coth{\frac{\pi(\Delta x - 2i\epsilon_{01})}{4\tau'}} 
    - \coth{\frac{\pi(\Delta x + 2i\epsilon_{01})}{4\tau'}}\nn \\
&- \tanh{\frac{\pi(-\Delta x + 2t_0)}{4\tau'}} 
    - \tanh{\frac{\pi(\Delta x + 2t_0)}{4\tau'}} 
    + 2\tanh{\frac{\pi(t_0 - i\epsilon_{01})}{2\tau'}} 
\big],
\end{align}
where $\epsilon_{01} := \epsilon_0 - \epsilon_1 > 0$, which come from the analytic continuation in (\ref{analytical_continuation}). It is crucial to retain the $i\epsilon$-prescription throughout the computation. As we will show below, it is closely related to the imaginary part of the timelike EE. A detailed discussion of the role of the $i\epsilon$-prescription is presented in the Appendix \ref{sec:d}.

Now, with this expression, one could evaluate the integrated term on the right-hand side of (\ref{relation_simpler}),  we have

\begin{align}
&\int_{-\Delta t}^{\Delta t} dy \, \partial_{t} S(t_0,y;t_0,0)\nn \\
&= \frac{c}{6} \!\left(
    2\log\!\left[ -\frac{\sinh{\frac{\pi(\Delta t - 2i\epsilon_{01})}{4\tau'}}}{\sinh{\frac{\pi(\Delta t + 2i\epsilon_{01})}{4\tau'}}} \right]
    + 2\log\!\left[ \frac{\cosh{\frac{\pi(\Delta t - 2t_0)}{4\tau'}}}{\cosh{\frac{\pi(\Delta t + 2t_0)}{4\tau'}}} \right]
    + \frac{\pi \Delta t}{\tau'} \tanh{\frac{\pi(t_0 - i\epsilon_{01})}{2\tau'}}
\right).
\end{align}
In the limit $\Delta t, t_0\gg \tau'$, we find
\begin{align}\label{integration_part}
\frac{1}{2}\int_{-\Delta t}^{\Delta t} dy \, \partial_{t} S(t_0,y;t_0,0)\simeq 
\begin{cases}
\dfrac{c\pi}{12\tau'}(\Delta t - 2t_0) + \dfrac{i c \pi}{6}, & t_0 < \Delta t/2, \\[6pt]
\dfrac{i c \pi}{6}, & t_0 > \Delta t/2.
\end{cases}
\end{align}
Note that the integrated term would contribute to the imaginary term of the timelike EE. This term also gives the contributions to the real EE, for the earlier time region $t_0<\frac{\Delta t}{2}$. For earlier times, the EE for a spacelike interval is growing, which can be understood in the quasiparticle picture. The temporal derivation gives the growing velocity of the EE. But at a later time $t_0>\frac{\Delta t}{2}$, the EE for a spacelike interval is saturated. Thus, the temporal derivative of EE would be vanishing, which leads to the vanishing of the real part of (\ref{integration_part}). However, the imaginary term would be constant at any time. As shown in \cite{Xu:2024yvf}, the imaginary term is closely related to the commutator of the twist operator and its temporal derivative. In fact, here the contribution of the imaginary term also comes from the commutator.

Using the results in Sec.~\ref{section_general_spacelike_global}, we can obtain
\bea\label{sum_part}
\frac{1}{2} \left[S(t_0,-\Delta t;\,t_0,0) + S(t_0,\Delta t;\,t_0,0)  \right]=
\begin{cases}
\frac{c}{3}\log \frac{2\tau'}{\pi}+\frac{c\pi\, t_0}{6\tau'}, & t_0<\frac{\Delta t}{2},\nn\\
\frac{c}{3}\log \frac{2\tau'}{\pi}+\frac{c\pi\, \Delta t}{12\tau'}, & t_0>\frac{\Delta t}{2}.
\end{cases}
\eea
For the earlier time $t_0< \frac{\Delta t}{2}$, the EE for the intervals $(-\Delta t,0)$ and $(0,\Delta t)$ is linearly growing. After time $t_0=\frac{\Delta t}{2}$, the EE is saturated and becomes a constant. Now we are ready to check the relation (\ref{relation_simpler}) by using Eq.~(\ref{integration_part}) and (\ref{sum_part}). This relation still holds for the global quench state. In the earlier time region $t_0<\frac{\Delta t}{2}$, the contribution from (\ref{sum_part}) is linearly growing, but the velocity of the EE growing is linearly decreasing, thus the time-dependent parts cancel with each other. Finally, for the entire time-evolution region, the total contributions are time-independent and depend only on the parameter $\Delta t$. This is consistent with the result that the timelike EE of $A(t_0,0;t_1,0)$ is time-independent, only depends on the time interval $\Delta t$. 

\subsection{Relation for locally excited state}

In this section, we would like to show that the relation (\ref{relation_simpler}) also holds for the locally excited state. Firstly, let us recall that for the locally excited state, the R\'enyi entropy for the general interval $A(t_0,x_0;t_1,x_1)$ can be decomposed as the vacuum contributions plus a correction, that is
\bea
S_n(t_0,x_0;t_1,x_1)=S_{n}^0(t_0,0;t_1,0)+\Delta S_n(t_0,x_0;t_1,x_1),
\eea
where $S_{n}^0(t_0,x_0;t_1,x_1)$ denotes the Rényi entropy in the vacuum state. For the timelike interval, $\Delta S_n$ is evaluated in Sec .~\ref {Section_general_timelike_local}. The results depend on four parameters: the position of the local operator $x$, the evolution time $t_0$, the space coordinate difference of the interval $L$, and the time coordinate difference of the interval $\Delta t$.

Firstly, we would like to note that the R\'enyi entropy in the vacuum state $S^0_n(t_0,0;t_1,0)$  is already satisfying the relation (\ref{relation_simpler}). We only need to check the relation 
\begin{eqnarray}\label{relation_simpler_local}
    \Delta S_n(t_0,0;t_1,0)= \frac{1}{2} \left[\Delta S_n(t_0,-\Delta t;\,t_0,0) + \Delta S_n(t_0,\Delta t;\,t_0,0)  \right]+\frac{1}{2}\int_{-\Delta t}^{\Delta t}dy \partial_t \Delta S_n(t_0,y;t_0,0).
\end{eqnarray}
Firstly, let us consider the integration term. In the locally excited state, the variance of R\'enyi entropy $\Delta S_n$ is only related to the quantum dimension of the primary operator $O_a$ in rational CFTs. But that does not mean the temporal derivative of $\Delta S_n$ would vanish. Since $\Delta S_n$ is a piecewise function, its time-dependence is not apparent.

Since the results depend on operator insertion position $x$, to simplify the discussion, we will consider  $x<-\Delta t$. To evaluate the temporal derivative $\partial_t \Delta S_n(t_0,y;t_0,0)$, we should consider $\Delta S_n(t,y;t',0)$ for the general spacelike interval $A(t,y;t',0)$, the make the evaluation $\partial_t \Delta S_n(t,y;t',0)|_{t\to t_0,t'\to t_0}$, where $y\in (-\Delta t,\Delta t)$. According to the results in Sec.~\ref{Section_general_spacelike_local} the EE for general spacelike interval $A(t,y;t',0)$, we can write down a compact formula for the R\'enyi entropy,
\bea
\Delta S_n(t,y;t',0)=
\begin{cases}
[\theta(t'+x)-\theta(t+x-y)]\log d_a,&y>0.\\
        [\theta(t+x-y)-\theta(t'+x)]\log d_a,&y<0,\\
\end{cases}
\label{eq:theta}
\eea
where $\theta$ is the Heaviside step function. See Appendix \ref{sec:e} for the derivation of the above formula. Now we can obtain
\bea
\partial_t \Delta S_n(t_0,y;t_0,0)=\partial_t \Delta S_n(t,y;t',0)|_{t\to t_0}
=\begin{cases}
    -\delta(t_0+x-y)\log d_a,&y>0,\\
    \delta(t_0+x-y)\log d_a,&y<0.
\end{cases}
\eea
Thus, the integration term is given by
\bea\label{local_integration}
&&\frac{1}{2}\int_{-\Delta t}^{\Delta t}dy \partial_t \Delta S_n(t_0,y;t_0,0)\nn\\
&&=\frac{1}{2}\int_{-\Delta t}^{0}dy \partial_t \Delta S_n(t_0,y;t_0,0)+\frac{1}{2}\int_{0}^{\Delta t}dy \partial_t \Delta S_n(t_0,y;t_0,0) \nn \\
&&=\begin{cases}
0,&t_0\in[0,-x-\Delta t)\\
\frac{1}{2}\log d_a, & t_0\in (-x-\Delta t,-x), \\
-\frac{1}{2}\log d_a, & t_0\in(-x,-x+\Delta t),\\
0,&t_0\in(-x+\Delta t,+\infty).
\end{cases}
\eea
Now let us consider $\Delta S_n(t_0,-\Delta t;\,t_0,0)$ and $\Delta S_n(t_0,\Delta t;\,t_0,0)$. Using the results in Sec.~\ref{Section_general_spacelike_local}, we have
\bea
\label{local_sum}
&& \frac{1}{2} \left[\Delta S_n(t_0,-\Delta t;\,t_0,0) + \Delta S_n(t_0,\Delta t;\,t_0,0)  \right]\nn\\
&&=
\begin{cases}
    \frac{1}{2}\log d_a,&t_0\in(-x-\Delta t,-x+\Delta t),\\
    0,&t_0\in[0,-x-\Delta t)\cup(-x+\Delta t,+\infty).
\end{cases}
\eea
By using the results in Sec.~\ref{Section_general_timelike_local}, the timelike EE associated with the interval $A(t_0,0;t_1,0)$ is given by
\bea\label{timelike_interval_local_t0t1}
\Delta S_n(t_0,0;t_1,0)=\begin{cases}
    \log d_a,&t_0\in(-x-\Delta t,-x),\\
     0,&t_0\in[0,-x-\Delta t)\cup(-x,+\infty),
\end{cases}.
\eea
 Now it is straightforwards to check the relation (\ref{relation_simpler_local}) is correct by using the results (\ref{timelike_interval_local_t0t1}) and (\ref{local_integration}) (\ref{local_sum}).

\section{Conclusion}\label{sec:6}

In this work, we have developed a systematic framework for computing EE  in $(1+1)$-dimensional CFTs for general spacetime intervals, with a particular focus on timelike intervals. By introducing the spacetime density matrix formalism, we have generalized the conventional density matrix to incorporate timelike correlations, ensuring that EE for both spacelike and timelike intervals is well-defined and computable, even in time-dependent states. This framework, utilizing the Schwinger-Keldysh path integral, extends the imaginary-time path-integral approach, enabling a unified treatment of static and dynamical systems.

We applied this framework to two widely studied models in CFTs: the global quench and the local quench models. In the global quench setup, we calculated EE for general spacelike intervals, revealing three distinct phases depending on the evolution time $t_0$ and the spatial and temporal separations, $\Delta x$ and $\Delta t$. While the behavior of EE for spacelike intervals aligns with previous studies, we highlighted significant differences when $\Delta t$ is nonzero, as discussed in Sec.~\ref{section_general_spacelike_global}. For timelike intervals, we found that EE remains constant over time, exhibiting qualitatively different behavior. We also generalized the quasiparticle picture, which has been applied to spacelike intervals, to timelike intervals, providing a coherent explanation for these dynamics, as detailed in Sec.~\ref{Section_quasiparticle_global}.

In the local quench model, we investigated the time evolution of the Rényi entropy and classified the entanglement contributions according to the spatial positions of the operator insertions. We found that the quasiparticle picture also applies in this case, identifying regimes in which entanglement increases due to partial intersections of quasiparticle worldlines with the entangling region, and other regimes in which no entanglement is generated. These results offer a comprehensive understanding of entanglement dynamics in arbitrary spacetime regions, enhancing our understanding of the causal structure of entanglement propagation in CFTs. 

However, we should emphasize that the quasiparticle picture for timelike intervals is essentially different from that for spacelike ones. In the spacelike case, the EE of an interval can be interpreted as arising from the entanglement between two quasiparticles, one located in the subsystem $A$ and the other in the complement on the same Cauchy surface, see Fig.~\ref{fig:4} for an illustration. In contrast, a timelike interval does not lie on a Cauchy surface, and therefore its complement cannot be defined in the usual sense. As a result, the EE associated with a timelike interval cannot be interpreted in the same way as in the spacelike case. The physical meaning of timelike EE thus remains unclear. We expect that the quasiparticle picture for the timelike interval considered in this paper can be more properly understood within the framework of the spacelike density matrix \cite{Milekhin:2025ycm, Guo:2025dtq}, which involves two Cauchy surfaces. We leave a detailed investigation of this issue for future work.

We also established the relation between time- and spacelike EE for generic states, previously known only for static states~\cite{Guo:2024lrr}. Our findings confirm that this relation holds in time-dependent states, establishing its universality. This non-trivial extension provides strong evidence that the time- and spacelike EE relation is a fundamental feature of CFTs. Although the physical meaning of timelike EE remains an open question, our results provide new insights into its connection with quasiparticles. In the quantum quench models studied, we interpret the real part of timelike EE as a quasiparticle contribution, analogous to spacelike intervals, while the imaginary part remains constant, resembling the vacuum state. This universal imaginary term is linked to the commutator of the twist operator and its temporal derivative~\cite{Xu:2024yvf}, and we show that quasiparticle excitations do not contribute to this term.

Although our analysis is based on two simple models, the general framework we developed can be extended to other models and theories. For example, the local quench model from \cite{Calabrese:2007mtj} can be adapted to compute timelike EE, and we plan to apply our methods to a broader range of models. Additionally, exploring timelike EE for theories with holographic duals presents an exciting direction. While a holographic dual for timelike EE is still under development, recent proposals~\cite{Doi:2022iyj,Heller:2024whi} and works~\cite{Li:2022tsv,Doi:2023zaf,Guo:2025pru,Nunez:2025ppd,Heller:2025kvp} have made significant progress. Studying the holographic dual of timelike EE in time-dependent states, especially in the global quench scenario where the holographic dual corresponds to a black hole with an end-of-world brane~\cite{Hartman:2013qma}, could provide valuable tests of these proposals. 

Moreover, in the context of dS/CFT \cite{Strominger:2001pn}, the bulk dynamics necessarily involves time evolution. As a consequence, the areas of extremal surfaces are expected to be complex-valued, sharing similar features with timelike EE \cite{Narayan:2022afv,Narayan:2023ebn,Nanda:2025tid}. This suggests an underlying non-Hermitian structure of the system \cite{Doi:2024nty}\cite{Harper:2025lav}. It would therefore be interesting to explore whether a spacetime density matrix for time-dependent states can provide a systematic framework to formalize and understand entanglement structures in the dS/CFT correspondence.

~\\

~\\

{\bf Acknowledgements}
We would like to thank Bin Chen, Jian-Xin Lu, Hong Lv,  Rong-Xin Miao, Jie-qiang Wu, Jin Xu, Run-qiu Yang, Jiaju Zhang, and Yang Zhou for useful discussions. WZG is supported by the Hubei Provincial Natural Science Foundation of China under Grant No.2025AFB557 and the Fundamental Research Support Program of the School of Physics, Huazhong University of Science and Technology. SH would like to appreciate the financial support from the Fundamental Research Funds for the Central Universities and the Natural Science Foundation of China (Grants No. 12475053, No. 12235016, and No. 12588101).

\appendix
\section{Two-point correlator of twist operators}
\label{sec:a}
We consider the two-point function of twist operators $\sigma_n$ and $\bar{\sigma}_n$ on a strip. 
Using the conformal map 
\begin{equation}
    w(z) = \frac{2\tau'}{\pi}\log z,
\end{equation}
which maps the UHP to a strip of width $2\tau'$, the correlator can be expressed in terms of the UHP correlator and the corresponding conformal factors.

\begin{align}
    & \langle \tilde{\sigma}_n(x_0,t_0)\, \sigma_n(x_1,t_1) \rangle_{\mathrm{strip}}\nn \\
    =&\left|\frac{\partial w_0}{\partial z}\right|^{-2h_n}\left|\frac{\partial w_1}{\partial z}\right|^{-2h_n}\langle \tilde{\sigma}_n(x_0,t_0)\, \sigma_n(x_1,t_1) \rangle_{UHP}\nonumber\\
    =&\left| \frac{\partial w_0}{\partial z} \right|^{-2h_n} \left| \frac{\partial w_1}{\partial z} \right|^{-2h_n} \left( \frac{z_{0\bar{1}} z_{1\bar{0}}}{z_{01} z_{\bar{0}\bar{1}} z_{0\bar{0}} z_{1\bar{1}}} \right)^{2h_n}F(\eta)\nonumber\\
    =& \left( \frac{\pi}{2\tau'} \right)^{4h_n}\left(  
    \frac{\left(1+e^{\frac{(\Delta x+2\Delta t+\Delta t)\pi}{2\tau'}}\right)\left(e^{\frac{(\Delta t+\Delta t)\pi}{\tau'}}+e^{\frac{(\Delta x+\Delta t)\pi}{2\tau'}}\right)}{\left(1+e^{\frac{\Delta t\pi}{\tau'}}\right)\left(e^{\frac{\Delta x\pi}{2\tau'}}-e^{\frac{t\pi}{2\tau'}}\right)\left(-1+e^{\frac{(\Delta x+\Delta t)\pi}{2\tau'}}\right)\left(1+e^{\frac{(\Delta t+\Delta t)\pi}{\tau'}}\right)}\right)^{2h_n} F(\eta)\nonumber\\
    =& \left( \frac{\pi}{2\tau'}\right)^{4h_n}\left(\frac{\cosh{\frac{(\Delta x+2t_0+\Delta t)\pi}{4\tau'}}\cosh{\frac{(\Delta x -2t_0-\Delta t)\pi}{4\tau'}}}{4\sinh{\frac{(\Delta x-\Delta t)\pi}{4\tau'}}\sinh{\frac{(\Delta x+\Delta t)\pi}{4\tau'}}\cosh{\frac{t_0\pi}{2\tau'}}\cosh{\frac{(t_0+\Delta t)\pi}{2\tau'}}} \right)^{2h_n} F(\eta).
\end{align}

%    =& \left( \frac{\pi}{2\tau'} \right)^{4h_n}\left(\frac{\cosh{\frac{(\Delta x+2t_0+\Delta t)\pi}{4\tau'}}\cosh{\frac{(\Delta x-2t_0-\Delta t)\pi}{4\tau'}}}{4\sinh{\frac{(\Delta x-\Delta t-2i\epsilon)\pi}{4\tau'}}\sinh{\frac{(\Delta x+\Delta t+2i\epsilon)\pi}{4\tau'}}\cosh{\frac{(t_0-i\epsilon)\pi}{2\tau'}}\cosh{\frac{(t_0+\Delta t+i\epsilon)\pi}{2\tau'}}} \right)^{2h_n} F(\eta)

\section{\texorpdfstring{The $n$-th R\'enyi entropy in the local quench model}
                         {The n-th Renyi Entropy in the local quench model}}

\label{sec:b}

To evaluate the change of the Rényi entropy after an excitation, we compare the $ n$-th Rényi entropy of the reduced density matrix $\rho^{EX}_A$ with that of the vacuum $\rho^{(0)}_A$. Using the replica method, the Rényi entropy is written in terms of partition functions on the $n$-sheeted Riemann surface. The difference$\Delta S^{(n)}_A$ can therefore be expressed as a ratio between the excited and vacuum partition functions. Introducing local operator insertions that generate the excited state, the partition functions are mapped to correlation functions of $2n$ operators on $\Sigma_n$ and the corresponding two-point function on $\Sigma_1$. This leads to the following expression:

\begin{align}
    \Delta S^{(n)}(\tau_0,0;\,\tau_1,L)=&S^{(n)}\!\left(\rho^{\mathrm{EX}}(\tau_0,0;\,\tau_1,L)\right)
       - S^{(n)}\!\left(\rho^{(0)}(\tau_0,0;\,\tau_1,L)\right) \notag\\
    =&\frac{1}{1-n}\log\frac{Z^{EX}_n}{(Z^{EX}_1)^n}-\frac{1}{1-n}\log\frac{Z_n}{(Z_1)^n}\nonumber\\
    =&\frac{1}{1-n}\log\frac{Z^{EX}_n(Z_1)^n}{Z_n(Z^{EX}_1)^n}\nonumber\\
    =&\frac{1}{1-n}\left(\log\frac{Z^{EX}_n}{Z_n}-\log\frac{(Z^{EX}_1)^n}{(Z_1)^n}\right)\nonumber\\
    =&\frac{1}{1-n}\log\langle O(w_1,\bar{w}_1)O(w_2,\bar{w}_2)\cdots O(w_{2n-1},\bar{w}_{2n-1})O(w_{2n},\bar{w}_{2n})\rangle_{\Sigma_n}\nn\\
    &-\frac{n}{1-n}\log\langle O(w_1,\bar{w}_1)O(w_2,\bar{w}_2)\rangle_{\Sigma_1} \nonumber\\
     =&\frac{1}{1-n}\log\frac{\langle O(w_1,\bar{w}_1)O(w_2,\bar{w}_2)\cdots O(w_{2n-1},\bar{w}_{2n-1})O(w_{2n},\bar{w}_{2n})\rangle_{\Sigma_n}}{\left(\langle O(w_1,\bar{w}_1)O(w_2,\bar{w}_2)\rangle_{\Sigma_1}\right)^n}.
\end{align}

\section{Details of the cross ratios}
\label{sec:c}
In this section we would like to show how to compute the holomorphic and anti-holomorphic cross-ratios $\eta$ and $\bar{\eta}$ in the local quench model. Introducing the short-hand variables $a, b, \alpha,$ and $\beta$ for the insertion coordinates and keeping a small $\epsilon$, we will compute the cross-ratios in the limit $\epsilon\to 0$. This leads to a piecewise behavior of $\eta$ and $\bar{\eta}$, taking the values $0$ or $1$ in different time intervals of $t_0$:

\begin{align}
    \eta &=\frac{z_{12}z_{34}}{z_{13}z_{24}}\nonumber\\
    &=\frac{-\left(\sqrt{\frac{-x-t_0-i\epsilon}{-x+L-t_0-\Delta t -i\epsilon}}
      -\sqrt{\frac{-x-t_0+i\epsilon}{-x+L-t_0-\Delta t +i\epsilon}}\right)^2}{
      4\sqrt{\frac{-x-t_0-i\epsilon}{-x+L-t_0-\Delta t -i\epsilon}}
       \sqrt{\frac{-x-t_0+i\epsilon}{-x+L-t_0-\Delta t +i\epsilon}}}
\end{align}
Here we define
\[a = -x - t_0, \qquad 
b = -x - t_0 + L - \Delta t ,\]
so that $\eta$ can be rewritten as
\begin{align}
    \eta &=
    \frac{-\left(\sqrt{\frac{a-i\epsilon}{b-i\epsilon}}
      -\sqrt{\frac{a+i\epsilon}{b+i\epsilon}}\right)^2}{
      4\sqrt{\frac{a-i\epsilon}{b-i\epsilon}}
       \sqrt{\frac{a+i\epsilon}{b+i\epsilon}}}\nonumber\\
    &=\frac{
      \sqrt{(a^2+\epsilon^2)(b^2+\epsilon^2)}
      -ab-\epsilon^2}{
      2\sqrt{(a^2+\epsilon^2)(b^2+\epsilon^2)}}\nonumber\\
    &\simeq 
    \frac{|ab|-ab}{2|ab|}
      +\frac{\epsilon^2(a^2+b^2)}{4a^2 b^2}\nonumber\\
    &\simeq
    \begin{cases}
        0, & ab>0,\nonumber\\
        1, & ab<0,
    \end{cases}\\
    &=
    \begin{cases}
        0, & t_0\in[0,-x)\cup(-x + L - \Delta t,+\infty),\\
        1, & t_0\in(-x,-x + L - \Delta t).
    \end{cases}
\end{align}

Similarly, we obtain

\begin{align}
    \bar{\eta}
    &=\frac{\bar{z}_{12}\bar{z}_{34}}{\bar{z}_{13}\bar{z}_{24}}\nonumber\\
     &=\frac{-\left(\sqrt{\frac{-x+t_0-i\epsilon}{-x+L+t_0-\Delta t -i\epsilon}}-\sqrt{\frac{-x+t_0+i\epsilon}{-x+L+t_0-\Delta t +i\epsilon}}\right)^2}{4\sqrt{\frac{-x+t_0-i\epsilon}{-x+L+t_0-\Delta t -i\epsilon}}\sqrt{\frac{-x+t_0+i\epsilon}{-x+L+t_0-\Delta t +i\epsilon}}}\nonumber\\
     &=\frac{-\left(\sqrt{\frac{\alpha-i\epsilon}{\beta-i\epsilon}}-\sqrt{\frac{\alpha+i\epsilon}{\beta +i\epsilon}}\right)^2}{4\sqrt{\frac{\alpha-i\epsilon}{\beta -i\epsilon}}\sqrt{\frac{\alpha+i\epsilon}{\beta +i\epsilon}}}\nonumber\\
    &=\frac{\sqrt{(\alpha^2+\epsilon^2)(\beta^2+\epsilon^2)}-\alpha\beta-\epsilon^2}
            {2\sqrt{(\alpha^2+\epsilon^2)(\beta^2+\epsilon^2)}}\nonumber\\
    &\simeq \frac{|\alpha\beta|-\alpha\beta}{2|\alpha\beta|}
      +\frac{\epsilon^2(\alpha^2+\beta^2)}{4\alpha^2\beta^2}\nonumber\\
    &=
    \begin{cases}
        0, & t_0\in[0, x - L - \Delta t)\cup(  x,+\infty),\\
        1, & t_0\in(x - L - \Delta t, x).
    \end{cases}
\end{align}
where $z_{ij}=z_i - z_j$,
$\alpha = -x + t_0,
\beta = -x + t_0 + L + \Delta t.$

Depending on the value of $t_0\ge0$, $(\eta,\bar{\eta})$ fall into four distinct cases, as summarized in Table~\ref{tab:eta}.

\begin{table}[ht]
    \centering
    \begin{tabular}{c|c|c}
    \hline
       $t_0$  & $\eta=0$ &  $\eta=1$ \\ 
    \hline
    $\bar{\eta}=0$ 
    &\makecell{$[0,-x)\cap[0,\, x - L - \Delta t)$ \\
\text{or}\\
$[0,-x)\cap(x,+\infty)$ \\
\text{or}\\
$[0,\, x - L - \Delta t)\cap(-x + L - \Delta t,+\infty)$ \\
\text{or}\\
$(-x + L - \Delta t,+\infty)\cap(x,+\infty)$
}    & \makecell{ $[0, x - L - \Delta t)\cap(-x,-x + L - \Delta t)$\\\text{or}\\$(-x,-x + L - \Delta t)\cap(  x,+\infty)$}\\
    \hline
    $\bar{\eta}=1$ &\makecell{$[0,-x)\cap(x - L - \Delta t, x)$\\\text{or}\\$(x - L - \Delta t, x)\cap( -x + L - \Delta t,+\infty)$} 
    &$(-x,-x + L - \Delta t)\cup(x - L - \Delta t, x)$\\  
    \hline
    \end{tabular}
    \caption{Different ranges of the insertion time $t_0 \ge 0$ give rise to the four possible values of the cross ratios $(\eta,\bar{\eta})$.}
    \label{tab:eta}
\end{table}

\section{\texorpdfstring{$i\epsilon$}{epsilon}-prescription and the imaginary part of timelike EE}
\label{sec:d}
We begin by noting that the infinitesimal regulators $\epsilon_{01}$ are essential in the complete timelike configuration. If they are omitted, the analytic continuation is no longer implemented correctly, and the timelike EE loses its imaginary part. To illustrate this, let us temporarily set $\epsilon_0 = \epsilon_1 = 0$.

The first time derivative of the EE for the interval $A(t_0,x_0;t_0,0)$ is
\begin{align}
\partial_{t} S(t_0,\Delta x;t_0,0) 
= \frac{c\pi}{24\tau'} \!\big[
    \coth{\frac{\pi(\Delta x - 2i\epsilon_{01})}{4\tau'}} 
    - \coth{\frac{\pi(\Delta x + 2i\epsilon_{01})}{4\tau'}} \nn \\
\quad
    - \tanh{\frac{\pi(-\Delta x + 2t_0)}{4\tau'}} 
    - \tanh{\frac{\pi(\Delta x + 2t_0)}{4\tau'}} 
    + 2\tanh{\frac{\pi(t_0 - i\epsilon_{01})}{2\tau'}} 
\big].
\end{align}
If we set $\epsilon_0 = \epsilon_1 = 0$, the contribution
\begin{align}
\big[
    \coth{\frac{\pi(\Delta x - 2i\epsilon_{01})}{4\tau'}} 
    - \coth{\frac{\pi(\Delta x + 2i\epsilon_{01})}{4\tau'}}
\big]
\end{align}
is effectively discarded. However, the integral of this term is precisely
\begin{align}
    &\int_{-\Delta t}^{\Delta t} dy\,
    \big[\coth{\frac{\pi(\Delta x - 2i\epsilon_{01})}{4\tau'}} 
    - \coth{\frac{\pi(\Delta x + 2i\epsilon_{01})}{4\tau'}}\big] \nn \\ 
    &= \log\!\left[ -\frac{\sinh{\frac{\pi(\Delta t - 2i\epsilon_{01})}{4\tau'}}}{\sinh{\frac{\pi(\Delta t + 2i\epsilon_{01})}{4\tau'}}} \right]= i\pi.
\end{align}
This contribution precisely reproduces the imaginary part of the timelike EE. Therefore, if the regulators $\epsilon_{01}$ are removed, this imaginary part would disappear.

\section{Derivation of Eq.(\ref{eq:theta})}
\label{sec:e}
We consider a local quench in a $(1+1)$-dimensional CFT and study the
EE of a spacelike interval
$A(t,y;t',0)$ with endpoints at $(t',0)$ and $(t,y)$, where $y>0$.
A local operator is inserted at position $x<0$ at time $t=0$, creating
a pair of entangled quasiparticles propagating along right- and left-moving
worldlines.

For $y>0$, at the initial time $t_0=0$, neither of the two quasiparticle
worldlines intersects the interval $A$, so the excess entanglement
entropy is $\Delta S_n=0$. At time $t_0=-x-t'$ the right-moving
quasiparticle worldline first enters the interval, and the entanglement
entropy jumps to $\Delta S_n=\log d_a$. At time $t_0=-x+t'$, this
worldline leaves the interval again, and the excess entanglement
entropy returns to zero.
The corresponding piecewise form of the excess Rényi entropy is
\begin{align}
    \Delta S_n(t,y;t',0)
    =
    \begin{cases}
        \log d_a, &t_0\in(0,\,-x-t')\cup(-x-t+y,\,+\infty),\\[4pt]
        0,        &t_0\in(-x-t',\,-x-t+y).
    \end{cases}
\end{align}
This can be compactly rewritten using Heaviside step functions as
\begin{align}
    \Delta S_n(t,y;t',0)
    =
    \big[\theta(t'+x)-\theta(t+x-y)\big]\log d_a .
\end{align}

For $y>0$, at the initial time $t_0=0$, neither of the two quasiparticle
worldlines intersects the interval $A$, so the excess entanglement
entropy is $\Delta S_n=0$. At time $t_0=-x-t+y$ the right-moving
quasiparticle worldline first enters the interval, and the entanglement
entropy jumps to $\Delta S_n=\log d_a$. At time $t_0=-x+t'$, the quasiparticle worldline leaves the interval again, and the excess EE returns to zero.
The corresponding piecewise form of the excess Rényi entropy is
\begin{align}
    \Delta S_n(t,y;t',0)
    =
    \begin{cases}
        \log d_a, &t_0\in(0,\,-x-t+y)\cup(-x-t',\,+\infty),\\[4pt]
        0,        &t_0\in(-x-t+y,\,-x-t').
    \end{cases}
\end{align}
which can be expressed in terms of step functions as
\begin{align}
    \Delta S_n(t,y;t',0)
    =
    \big[\theta(t+x-y)-\theta(t'+x)\big]\log d_a .
\end{align}

Combining both cases, the final compact expression is
\begin{align}
\Delta S_n(t,y;t',0)
=
\begin{cases}
\big[\theta(t'+x)-\theta(t+x-y)\big]\log d_a, & y>0,\\[4pt]
\big[\theta(t+x-y)-\theta(t'+x)\big]\log d_a, & y<0.
\end{cases}
\end{align}

\end{document}